\documentclass[prd,twocolumn,showpacs,amsmath,amssymb,flushbottom,nofootinbib]{revtex4-1}

\usepackage{graphicx}
\usepackage{epstopdf}
\usepackage{tabularx}
\usepackage{multirow}

\usepackage[colorlinks,citecolor=darkblue,linkcolor=darkblue,urlcolor=darkred]{hyperref}

\usepackage{xcolor}
\xdefinecolor{darkred}{rgb}{0.35,0.,0.}
\xdefinecolor{darkgreen}{rgb}{0.,0.35,0.}
\xdefinecolor{darkblue}{rgb}{0.,0.,0.35}

\usepackage{times}

\usepackage[pagewise]{lineno}

\usepackage{relsize}
\def\babar{\mbox{\slshape B\kern-0.1em{\smaller A}\kern-0.1em
    B\kern-0.1em{\smaller A\kern-0.2em R}}}

\newcommand{\ST}{\rule[0.75em]{0pt}{0.75em}}

\begin{document}

\title{\boldmath
Analysis of $D^+\to\bar K^0e^+\nu_e$ and $D^+\to\pi^0e^+\nu_e$ Semileptonic Decays
}
\author{
\small{
\begin{center}
M.~Ablikim$^{1}$, M.~N.~Achasov$^{9,d}$, S.~Ahmed$^{14}$, X.~C.~Ai$^{1}$, O.~Albayrak$^{5}$, M.~Albrecht$^{4}$, D.~J.~Ambrose$^{45}$, A.~Amoroso$^{50A,50C}$, F.~F.~An$^{1}$, Q.~An$^{47,38}$
% ...
\\
\vspace{0.2cm}
(BESIII Collaboration)\\
\vspace{0.2cm} {\it
$^{1}$ Institute of High Energy Physics, Beijing 100049, People's Republic of China\\
% ...
$^{4}$ Bochum Ruhr-University, D-44780 Bochum, Germany\\
$^{5}$ Carnegie Mellon University, Pittsburgh, Pennsylvania 15213, USA\\
% ...
$^{9}$ G.I. Budker Institute of Nuclear Physics SB RAS (BINP), Novosibirsk 630090, Russia\\
% ...
$^{14}$ Helmholtz Institute Mainz, Johann-Joachim-Becher-Weg 45, D-55099 Mainz, Germany\\
% ...
$^{38}$ State Key Laboratory of Particle Detection and Electronics, Beijing 100049, Hefei 230026, People's Republic of China\\
% ...
$^{45}$ University of Rochester, Rochester, New York 14627, USA\\
% ...
$^{47}$ University of Science and Technology of China, Hefei 230026, People's Republic of China\\
% ...
$^{50}$ (A)University of Turin, I-10125, Turin, Italy; (B)University of Eastern Piedmont, I-15121, Alessandria, Italy; (C)INFN, I-10125, Turin, Italy\\
% ...
$^{d}$ Also at the Novosibirsk State University, Novosibirsk, 630090, Russia\\
% ...
}\end{center}
\vspace{0.4cm}
}
}
\noaffiliation
\date{\today}

\begin{abstract}
Using 2.93~fb$^{-1}$ of data taken at 3.773 GeV with the BESIII detector operated
at the BEPCII collider, we study the semileptonic decays $D^+ \to \bar K^0e^+\nu_e$ and
$D^+ \to \pi^0 e^+\nu_e$. We measure the absolute decay branching fractions
$\mathcal B(D^+ \to \bar K^0e^+\nu_e)=(8.60\pm0.06\pm 0.15)\times10^{-2}$ and
$\mathcal B(D^+ \to \pi^0e^+\nu_e)=(3.63\pm0.08\pm0.05)\times10^{-3}$,
where the first uncertainties are statistical and the second systematic.
We also measure the differential decay rates
and study the form factors of these two decays.
With the values of $|V_{cs}|$ and $|V_{cd}|$ from Particle Data Group fits assuming CKM unitarity,
we obtain the values of the form factors at $q^2=0$,
$f^K_+(0) = 0.725\pm0.004\pm 0.012$ and
$f^{\pi}_+(0) = 0.622\pm0.012\pm 0.003$.
Taking input from recent lattice QCD calculations of these form factors,
we determine values of the CKM matrix elements
$|V_{cs}|=0.944 \pm 0.005 \pm 0.015 \pm 0.024$ and
$|V_{cd}|=0.210 \pm 0.004 \pm 0.001 \pm 0.009$,
where the third uncertainties are theoretical.
\end{abstract}



\maketitle

\section{Introduction}
\label{sec:intro}

In the Standard Model (SM) of particle physics, the mixing between the
quark flavours in the weak interaction is parameterized by the Cabibbo-Kobayashi-Maskawa (CKM) matrix, which
is a $3\times3$ unitary matrix.
Since the CKM matrix elements are fundamental parameters of the SM,
precise determinations of these elements are very important for tests
of the SM and searches for New Physics (NP) beyond the SM.

Since the effects of strong and weak interactions can be well separated in
semileptonic $D$ decays, these decays are excellent processes from which we can determine the magnitude of the CKM matrix element $V_{cs(d)}$.
In the SM, neglecting the lepton mass, the differential decay rate for $D^+\to P e^+\nu_e$
($P= \bar K^0$ or $\pi^0$) is given by~\cite{Z_Phys_C_46_93}
\begin{linenomath*}
\begin{equation}\label{eq:dG_dq2}
    \frac{d\Gamma}{dq^2} = X\frac{G_F^2}{24\pi^3}|V_{cs(d)}|^2 p^3|f_+(q^2)|^2,
\end{equation}
\end{linenomath*}
where
$G_F$ is the Fermi constant,
$V_{cs(d)}$ is the corresponding CKM matrix element,
$p$ is the momentum of the meson $P$ in the rest frame of the $D$ meson,
$q^2$ is the squared four momentum transfer,
\emph{i.e.}, the invariant mass of the lepton and neutrino system,
and $f_+(q^2)$ is the form factor which parameterizes the effect of the strong interaction.
In Eq.~(\ref{eq:dG_dq2}), $X$ is a multiplicative factor due to isospin,
which equals to 1 for the decay $D^+\to\bar K^0e^+\nu_e$ and $1/2$ for the decay $D^+\to\pi^0e^+\nu_e$.

In this article, we report the experimental study of $D^+\to\bar K^0e^+\nu_e$
and $D^+\to\pi^0e^+\nu_e$ decays
using a 2.93~fb$^{-1}$~\cite{lum} data set collected at a center-of-mass energy of
$\sqrt{s}=3.773$~GeV with the BESIII detector operated at the BEPCII collider.
Throughout this paper, the inclusion of charge conjugate
channels is implied.

The paper is structured as follows.
We briefly describe the BESIII detector and the Monte Carlo (MC) simulation in Sec.~\ref{sec:bes3}.
The event selection is presented in Sec.~\ref{sec:sel}.
The measurements of the absolute branching fractions and the differential decay rates
are described in Sec.~\ref{sec:BF} and \ref{sec:DR}, respectively.
In Sec.~\ref{sec:FF} we discuss the determination of form factors from the
measurements of decay rates, and finally, in Sec.~\ref{sec:CKM}, we present
the determination of the magnitudes of the CKM matrix elements $V_{cs}$ and $V_{cd}$.
A brief summary is given in Sec.~\ref{sec:sum}.

\section{BESIII detector}
\label{sec:bes3}

The BESIII detector is a cylindrical detector with a
solid-angle coverage of $93\%$ of $4\pi$,
designed for the study of hadron spectroscopy and $\tau$-charm physics.
The BESIII detector is described in detail in Ref.~\cite{bes3}.
Detector components particularly relevant for this work are
(1) the main drift chamber (MDC) with 43 layers surrounding the beam pipe,
which performs precise determination of charged particle
trajectories and provides a measurement of the specific ionization energy loss ($dE/dx$);
(2) a time-of-flight system (TOF) made of plastic scintillator counters, which are located outside of
the MDC and provide additional charged particle identification
information;
and (3) the electromagnetic calorimeter (EMC) consisting of 6240 CsI(Tl) crystals,
used to measure the energy of photons and to identify electrons.

A {\sc geant4}-based~\cite{geant4} MC simulation software~\cite{BOOST},
which contains the detector geometry description and the detector response,
is used to optimize the event selection criteria, study possible backgrounds,
and determine the reconstruction efficiencies.
The production of the $\psi(3770)$,
initial state radiation production of $\psi(3686)$ and $J/\psi$,
as well as the continuum processes of
$e^+e^-\to\tau^+\tau^-$ and $e^+e^-\to q\bar q$ ($q=u,d,s$) are simulated
by the MC event generator {\sc kkmc}~\cite{kkmc};
the known decay modes are generated by {\sc evtgen}~\cite{besevtgen}
with the branching fractions set to the world average values from the
Particle Data Group (PDG)~\cite{pdg2016};
while the remaining unknown decay modes are modeled by
{\sc lundcharm}~\cite{lundcharm}.
We also generate signal MC events consisting of
$\psi(3770)\to D^+D^-$ events in which the $D^-$ meson
decays to all possible final states
and the $D^+$ meson decays to a hadronic
or a semileptonic decay final state being investigated.
In the generation of signal MC events,
the semileptonic decays $D^+\to\bar K^0e^+\nu_e$ and $D^+\to\pi^0e^+\nu_e$
are modeled by the the modified pole parametrization (see Sec.~\ref{sec:form_facor}).

\section{Event reconstruction}
\label{sec:sel}

The center-of-mass energy of 3.773~GeV corresponds to
the peak of the $\psi(3770)$ resonance, which decays predominantly
into $D\bar D$ ($D^0\bar D^0$ or $D^+D^-$) meson pairs.
In events where a $D^-$ meson is fully reconstructed,
the remaining particles must all be decay
products of the accompanying $D^+$ meson. In the following, the
reconstructed meson is called ``tagged $D^-$'' or ``$D^-$ tag''.
In a tagged $D^-$ data sample, the recoiling $D^+$ decays to $\bar K^0e^+\nu_e$
or $\pi^0e^+\nu_e$ can be cleanly isolated and used to measure the branching fraction and
differential decay rates.

\subsection{Selection of $D^-$ tags}
\label{sec:sel:tag}

We reconstruct $D^-$ tags in the following nine hadronic modes:
$D^-\to K^+\pi^-\pi^-$, $D^-\to K^0_S\pi^-$, $D^-\to K^0_S K^-$, $D^-\to K^+K^-\pi^-$,
$D^-\to K^+\pi^-\pi^-\pi^0$, $D^-\to\pi^+\pi^-\pi^-$~\footnote{
We veto the $K^0_S\pi^-$ candidates when a $\pi^+\pi^-$ invariant mass
falls within the $K_S^0$ mass window.
}, $D^-\to K^0_S\pi^-\pi^0$,
$D^-\to K^+\pi^-\pi^-\pi^-\pi^+$, and $D^-\to K^0_S\pi^-\pi^-\pi^+$.
The selection criteria of $D^-$ tags used here are the same as those described in
Ref.~\cite{BESIII_Dptomunu}.
Tagged $D^-$ mesons are identified by their beam-energy-constrained mass
$M_{\rm BC} \equiv \sqrt {E_{\rm beam}^2/c^4-|\vec {p}_{\rm tag}|^2/c^2}$,
where $E_{\rm beam}$ is the beam energy,
and $\vec p_{\rm tag}$ is the measured 3-momentum of the tag candidate~\footnote{
In this analysis, all four-momentum vectors measured
in the laboratory frame are boosted to the $e^+e^-$ center-of-mass
frame.
}.
We also use the variable $\Delta E\equiv E_{\rm tag}-E_{\rm beam}$,
where $E_{\rm tag}$ is the measured energy of the tag candidate,
to select the $D^-$ tags.
Each tag candidate is subjected to a tag mode-dependent $\Delta E$ requirement as shown in Table~\ref{tab:Dtag}.
If there are multiple candidates per tag mode for an event,
the one with the smallest value of $|\Delta E|$ is retained.

\begin{table*}
\caption{
The $\Delta E$ requirements, the $M_{\rm BC}$ signal regions,
the yields of the $D^-$ tags ($N_{\rm tag}$) reconstructed in data,
and the reconstruction efficiency ($\varepsilon_{\rm tag}$) of $D^-$ tags.
The uncertainties are statistical only.
}
\label{tab:Dtag}
\centering
\begin{ruledtabular}
\begin{tabular}{lcccc}
 Tag mode                    & $\Delta E$ (MeV) & $M_{\rm BC}$ (GeV$/c^2$) &  $N_{\rm tag}$ & $\varepsilon_{\rm tag}$ (\%) \\
\hline
$D^-\to K^+\pi^-\pi^-$              &  $(-45,45)$ & $(1.8640,1.8770)$ &  $806830 \pm 1070$  & $51.8\pm0.1$ \\
$D^-\to K^0_S\pi^-$                 &  $(-45,45)$ & $(1.8640,1.8770)$ &  $102755 \pm 372 $  & $56.2\pm0.2$ \\
$D^-\to K^0_S K^-$                  &  $(-45,45)$ & $(1.8650,1.8770)$ &  $ 19566 \pm 185 $  & $52.1\pm0.5$ \\
$D^-\to K^+K^-\pi^-$                &  $(-50,50)$ & $(1.8650,1.8780)$ &  $ 68216 \pm 966 $  & $41.2\pm0.3$ \\
$D^-\to K^+\pi^-\pi^-\pi^0$         &  $(-78,78)$ & $(1.8620,1.8790)$ &  $271571 \pm 2367$  & $27.3\pm0.1$ \\
$D^-\to \pi^+\pi^-\pi^-$            &  $(-45,45)$ & $(1.8640,1.8770)$ &  $ 32150 \pm 371 $  & $56.9\pm0.7$ \\
$D^-\to K^0_S\pi^-\pi^0$            &  $(-75,75)$ & $(1.8640,1.8790)$ &  $245303 \pm 1273$  & $31.3\pm0.1$ \\
$D^-\to K^+\pi^-\pi^-\pi^-\pi^+$    &  $(-52,52)$ & $(1.8630,1.8775)$ &  $ 30923 \pm 733 $  & $22.1\pm0.2$ \\
$D^-\to K^0_S\pi^-\pi^-\pi^+$       &  $(-50,50)$ & $(1.8640,1.8770)$ &  $125740 \pm 1203$  & $33.0\pm0.2$ \\
\hline
Sum                          &             &                   &  $1703054 \pm 3405$  \\
\end{tabular}
\end{ruledtabular}
\end{table*}

\begin{figure*}
\centering
\includegraphics[width=0.72\textwidth]{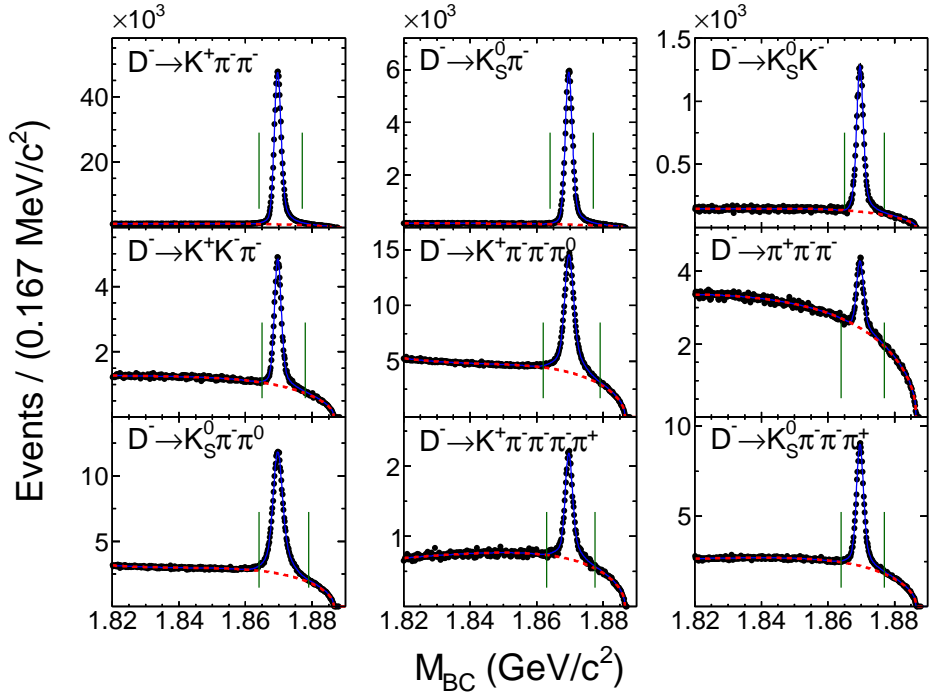}
\caption{
Fits (solid lines) to the $M_{\rm BC}$ distributions (points with error bars)
in data for nine $D^-$ tag modes.
The two vertical lines show the tagged $D^-$ mass regions.
}
\label{fig:Mbc}
\end{figure*}

The $M_{\rm BC}$ distributions for the nine $D^-$ tag modes are shown in Fig.~\ref{fig:Mbc}.
A binned extended maximum likelihood fit is used to determine the number of tagged $D^-$ events
for each of the nine modes.
We use the MC simulated signal shape
convolved with a double-Gaussian resolution function to represent
the beam-energy-constrained mass signal for the $D^-$ daughter particles, and
an ARGUS function \cite{Albrecht-1990am} multiplied by a third-order polynomial~\cite{BESII_D0toPenu, BESIII_D0toPenu}
to describe the background shape for the $M_{\rm BC}$ distributions.
In the fits all parameters of the double-Gaussian function,
the ARGUS function, and the polynomial function are left free.
The solid lines in Fig.~\ref{fig:Mbc} show the best fits, while the dashed
lines show the fitted background shapes.
The numbers of the $D^-$ tags ($N_{\rm tag}$)
within the $M_{\rm BC}$ signal regions
given by the two vertical lines in Fig.~\ref{fig:Mbc}
are summarized in Table~\ref{tab:Dtag}.
In total, we find $1703054 \pm 3405$ single $D^-$ tags reconstructed in data.
The reconstruction efficiencies of the single $D^-$ tags, $\epsilon_{\rm tag}$,
as determined with the MC simulation, are shown in Table~\ref{tab:Dtag}.

\subsection{Reconstruction of semileptonic decays}
\label{sec:sel:sl}

Candidates for semileptonic decays are selected from the remaining tracks in the system recoiling against the $D^-$ tags.
The $dE/dx$, TOF and EMC measurements (deposited energy and shape of the electromagnetic shower) are combined to form confidence levels for
the $e$ hypothesis ($CL_e$), the $\pi$ hypothesis ($CL_\pi$), and the $K$ hypothesis ($CL_K$).
Positron candidates are required to have $CL_e$ greater than 0.1\% and to satisfy $CL_e/(CL_e + CL_{\pi} + CL_K)>0.8$.
In addition, we include the 4-momenta of near-by photons
within $5^\circ$ of
the direction of the positron momentum
to partially account for final-state-radiation energy losses (FSR recovery).
The neutral kaon candidates are built from pairs of oppositely charged tracks that
are assumed to be pions.
For each pair of charged tracks, a vertex fit is performed and
the resulting track parameters are used to calculate the invariant mass, $M(\pi^+\pi^-)$.
If $M(\pi^+\pi^-)$ is in the range (0.484, 0.512)~GeV$/c^2$, the $\pi^+\pi^-$
pair is treated as a $K_S^0$ candidate and is used for further analysis.
The neutral pion candidates are reconstructed via the $\pi^0\to\gamma\gamma$ decays.
For the photon selection, we require
the energy of the shower deposited
in the barrel (end-cap) EMC greater than
25 (50)~MeV and
the shower time be within 700~ns of the event start time.
In addition, the angle between the photon and the
nearest charged track is required to be greater than $10^\circ$.
We accept the pair of photons as a $\pi^0$ candidate
if the invariant mass of the two photons, $M(\gamma\gamma)$,
is in the range (0.110, 0.150)~GeV$/c^2$.
A 1-Constraint (1-C) kinematic fit is then performed to constrain
 $M(\gamma\gamma)$ to the $\pi^0$ nominal mass,
and the resulting 4-momentum of the candidate $\pi^0$ is used for further analysis.

We reconstruct the $D^+\to \bar K^0 e^+\nu _e$ decay by requiring exactly three
additional charged tracks in the rest of the event. One track with charge opposite to
that of the $D^-$ tag is identified as
a positron using the criteria mentioned above,
while the other two oppositely charged
tracks form a $K_S^0$ candidate.
For the selection of the $D^+\to \pi^0 e^+\nu _e$ decay, we require that
there is only one additional charged track consistent with the positron
identification criteria and at least two photons that are used to form a $\pi^0$
candidate in the rest of the event. If there are
multiple $\pi^0$ candidates,
the one with the minimum $\chi^2$ from the 1-C kinematic fit is retained.
In order to additionally suppress background due to wrongly reconstructed or background photons,
the semileptonic candidate is further required to have
the maximum energy of any of the unused photons, $E_{\gamma,\rm max}$, less than 300~MeV.

Since the neutrino is undetected,
the kinematic variable $U_{\rm miss}\equiv E_{\rm miss}- c|\vec p_{\rm miss}|$
is used to obtain the information about the missing neutrino,
where $E_{\rm miss}$ and $\vec p_{\rm miss}$ are, respectively, the total missing energy
and momentum in the event.
The missing energy is computed from $E_{\rm miss} = E_{\rm beam} - E_{P} - E_{e^+}$,
where
$E_{P}$ and $E_{e^+}$ are the measured energies of the pseudoscalar meson and the positron, respectively.
The missing momentum $\vec p_{\rm miss}$ is given by $\vec p_{\rm miss} = \vec p_{D^+} - \vec p_{P} - \vec p_{e^+}$,
where $\vec p_{D^+}$, $\vec p_{P}$ and $\vec p_{e^+}$ are
the 3-momenta of the $D^+$ meson, the pseudoscalar meson and the positron, respectively.
The 3-momentum of the $D^+$ meson is taken as
$\vec p_{D^+} = - \hat p_{\rm tag} \sqrt{(E_{\rm beam}/c)^2 - (m_{D^+}c)^2}$,
where $\hat p_{\rm tag}$ is the direction of the momentum of the single $D^-$ tag,
and $m_{D^+}$ is the $D^+$ mass.
If the daughter particles from a semileptonic decay are correctly identified,
$U_{\rm miss}$ is near zero, since only one neutrino is missing.

\begin{figure}
\centerline{
\includegraphics[width=0.24\textwidth]{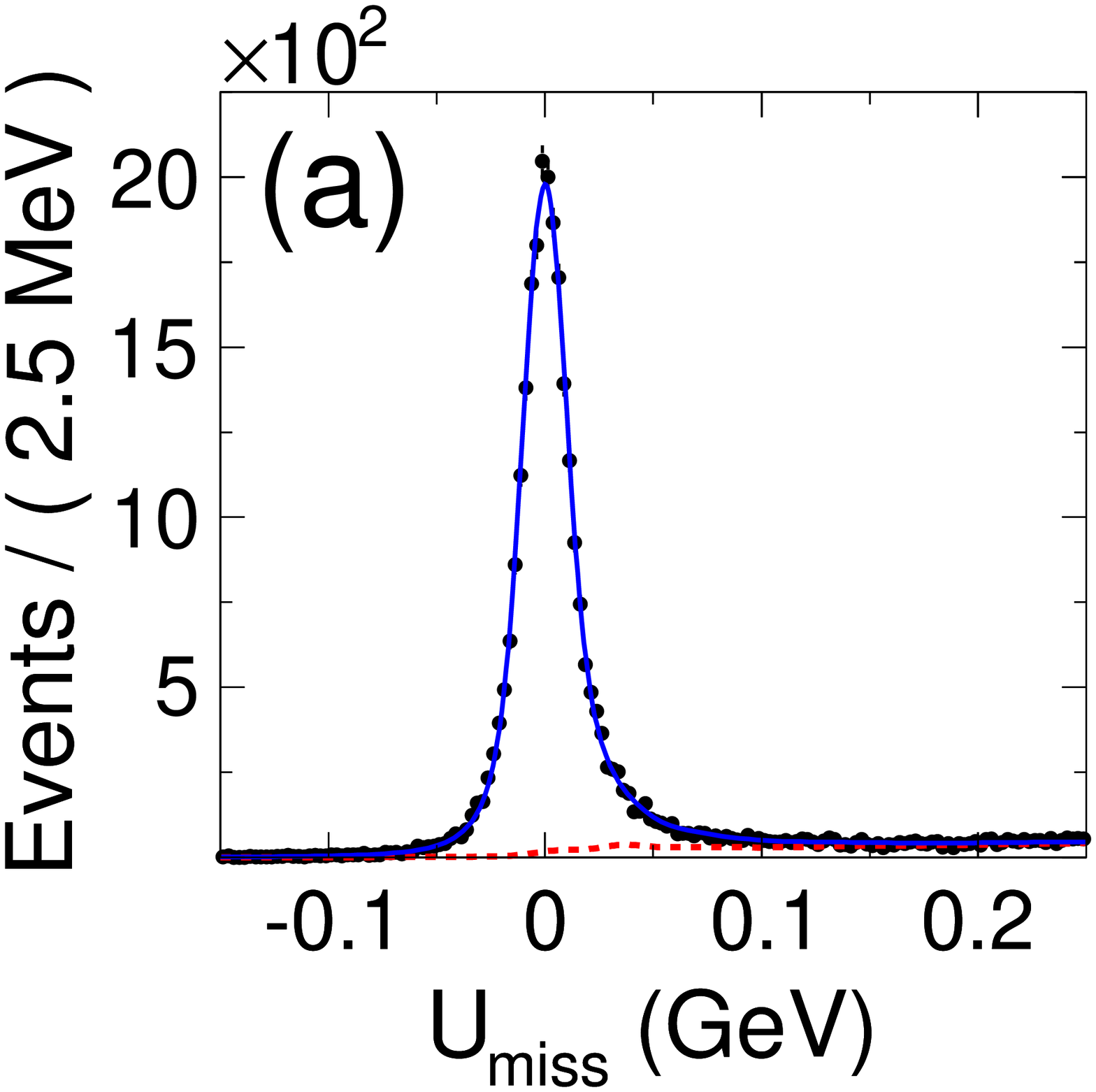}
\includegraphics[width=0.24\textwidth]{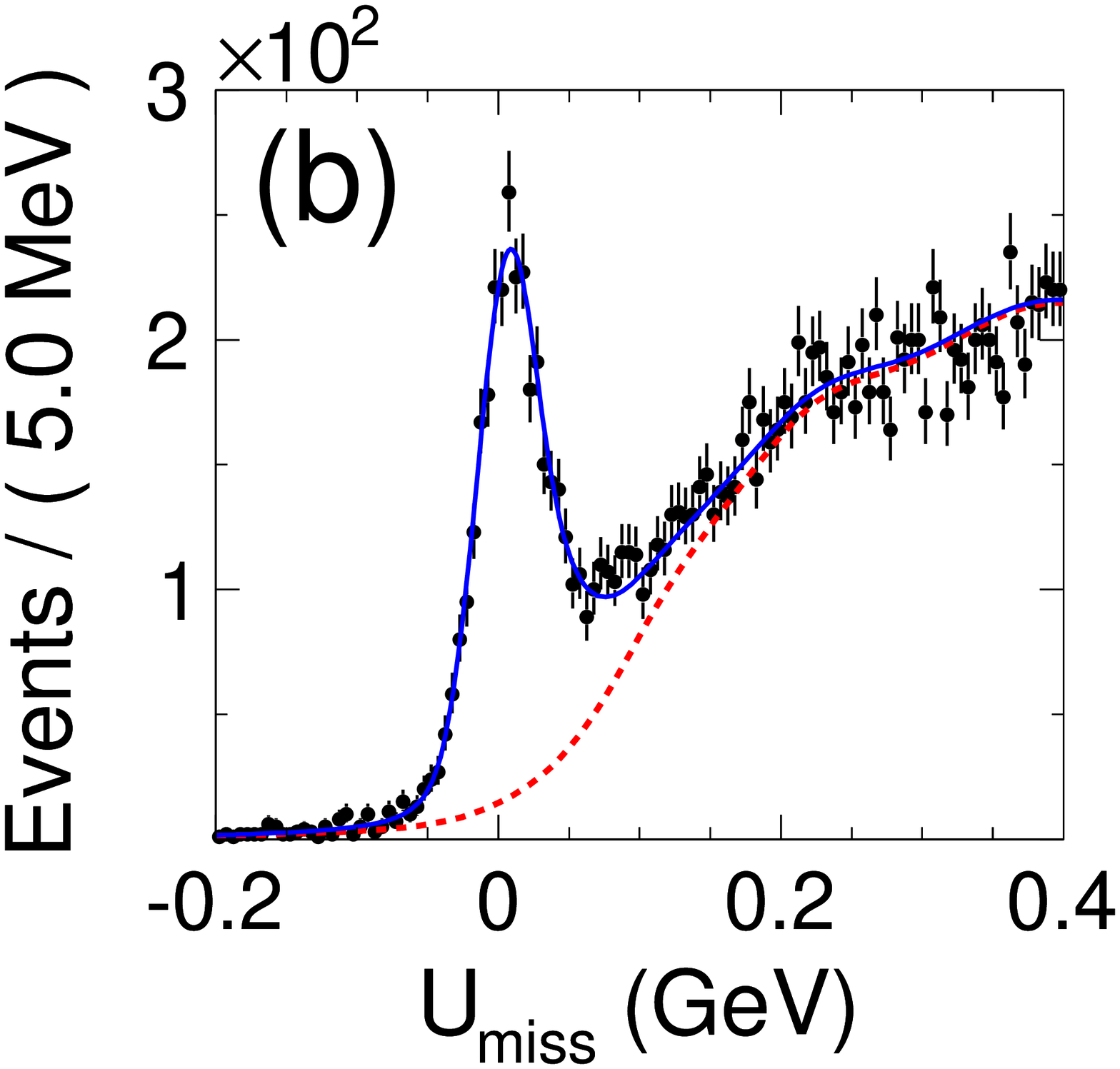}
}
\caption{
Distributions of $U_{\rm miss}$ for the selected (a) $D^+\to\bar K^0e^+\nu_e$ and (b) $D^+\to\pi^0e^+\nu_e$ candidates (points with error bars) with fit projections overlaid (solid lines).  The dashed curves show the background determined by the fit.}
\label{fig:Umiss}
\end{figure}

Figure~\ref{fig:Umiss} shows the $U_{\rm miss}$ distributions for the semileptonic candidates,
where the potential backgrounds arise from the $D\bar D$ processes other than signal,
$\psi(3770)\to$ non-$D\bar D$ decays, $e^+e^-\to\tau^+\tau^-$,
continuum light hadron production, initial state radiation return to $J/\psi$ and $\psi(3686)$.
The background for $D^+\to\bar K^0e^+\nu_e$ is dominated by
$D^+\to\bar K^*(892)^0e^+\nu_e$ and $D^+\to\bar K^0\mu^+\nu_\mu$.
For $D^+\to\pi^0e^+\nu_e$, the background is mainly from
$D^+\to K_L^0e^+\nu_e$ and $D^+\to K_S^0(\pi^0\pi^0)e^+\nu_e$.

Following the same procedure described in Ref.~\cite{BESIII_D0toPenu},
we perform a binned extended maximum likelihood fit to the $U_{\rm miss}$ distribution for each
channel to separate the signal from the background component.
The signal shape is constructed from a convolution of a MC determined distribution and a Gaussian function that accounts for the difference of the
$U_{\rm miss}$ resolutions between data and MC simulation.
The background shape is formed from MC simulation.
From the fits shown as the overlaid curves in Fig.~\ref{fig:Umiss},
we obtain the yields of the observed signal events to be
$N_{\rm obs}(D^+\to\bar K^0 e^+\nu_e)=26008\pm168$ and
$N_{\rm obs}(D^+\to\pi^0 e^+\nu_e)=3402\pm70$, respectively.

\begin{figure}
\centerline{
\includegraphics[width=0.48\textwidth]{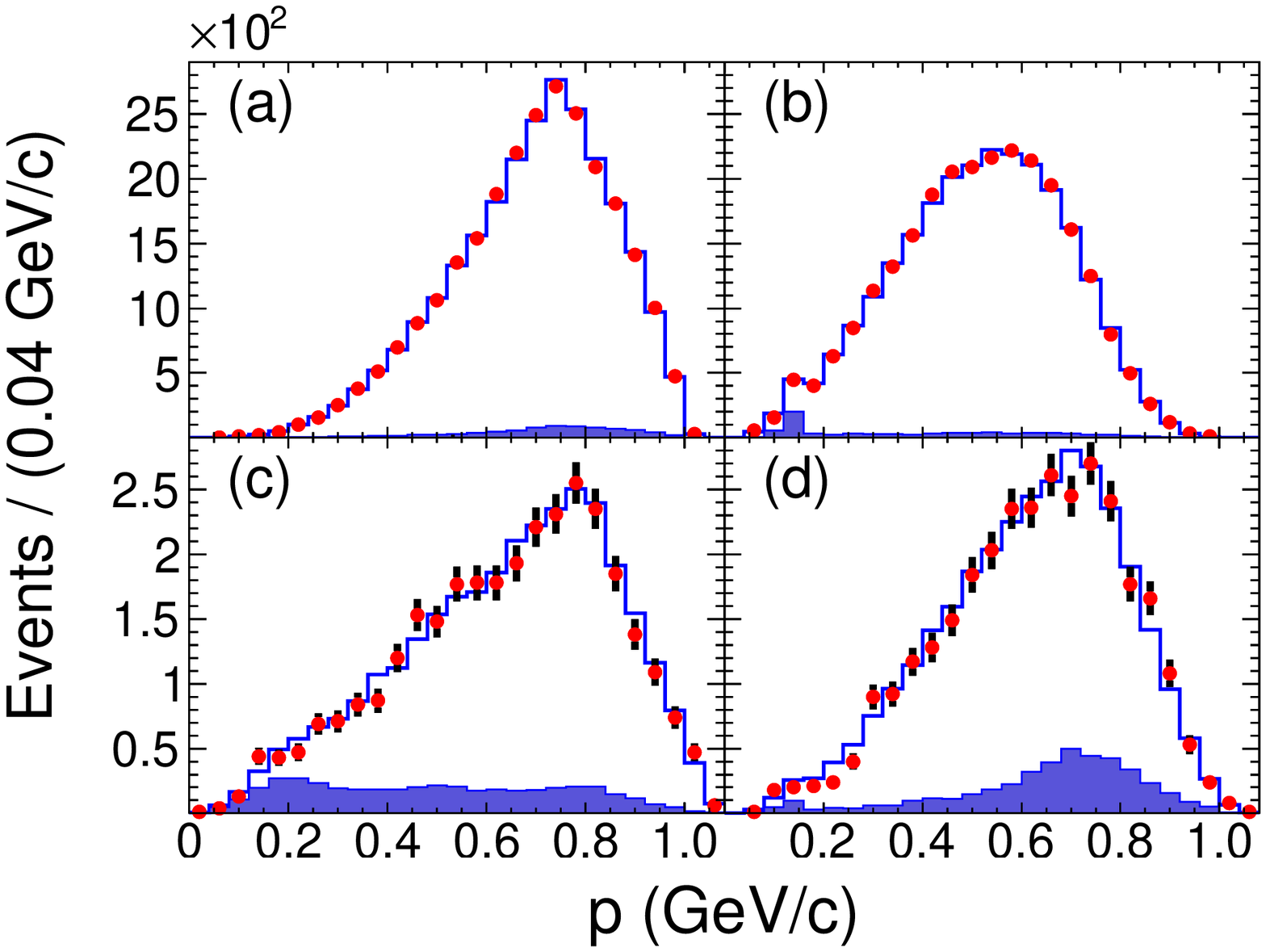}
}
\caption{
Momentum distributions of selected events (with $|U_{\rm miss}|<60$~MeV)
for (a) $\bar K^0$, (b)$e^+$ from $D^+\to\bar K^0 e^+\nu_e$,
(c) $\pi^0$, and (d) $e^+$ from $D^+\to\pi^0 e^+\nu_e$.
The points with error bars represent data,
the (blue) open histograms are MC simulated signal plus background,
the shaded histograms are MC simulated background only.
}
\label{fig:p}
\end{figure}

To check the quality of the MC simulation,
we examine the distributions of the reconstructed kinematic variables.
Figure~\ref{fig:p} shows the comparisons of the momentum distributions of
data and MC simulation.

\section{Branching fraction measurements}
\label{sec:BF}

\subsection{Determinations of branching fractions}

The branching fraction of the semileptonic decay $D^+\to Pe^+\nu_e$ is
obtained from
\begin{linenomath*}
\begin{equation}\label{eq:BF}
    \mathcal B(D^+\to Pe^+\nu_e) = \frac{N_{\rm obs}(D^+\to Pe^+\nu_e)}{N_{\rm tag}\,\varepsilon(D^+\to Pe^+\nu_e)},
\end{equation}
\end{linenomath*}
where $N_{\rm tag}$ is the number of $D^-$ tags (see Sec.~\ref{sec:sel:tag}),
$N_{\rm obs}(D^+\to Pe^+\nu_e)$ is the number of observed $D^+\to Pe^+\nu_e$ decays
within the $D^-$ tags (see Sec.~\ref{sec:sel:sl}), and $\varepsilon(D^+\to Pe^+\nu_e)$
is the reconstruction efficiency.
Here the $D^+\to\bar K^0 e^+\nu_e$ efficiency includes
the $K^0_S$ fraction of the $\bar K^0$ and $K^0_S\to\pi^+\pi^-$ branching fraction,
the $D^+\to\pi^0 e^+\nu_e$ efficiency includes
the $\pi^0\to\gamma\gamma$ branching fraction~\cite{pdg2016}.

Due to the difference in the multiplicity, the reconstruction efficiency varies slightly with the tag mode.
For each tag mode $i$, the reconstruction efficiency is given by
$\varepsilon^i=\varepsilon^i_{\rm tag,SL}/\varepsilon^i_{\rm tag}$,
where the efficiency for simultaneously finding
the $D^+\to Pe^+\nu_e$ semileptonic decay and the $D^-$ meson tagged with
mode $i$, $\varepsilon^i_{\rm tag,SL}$, is determined using the signal MC sample,
and $\varepsilon^i_{\rm tag}$ is the corresponding tag efficiency shown in
Table~\ref{tab:Dtag}.
These efficiencies are listed in Table~\ref{tab:eff}.
The reconstruction efficiency for each tag mode is then weighted according to the
corresponding tag yield in data to obtain the average reconstruction efficiency,
$\bar\varepsilon=\sum_i(N_{\rm tag}^i\varepsilon^i)/N_{\rm tag}$,
as listed in the last row in Table~\ref{tab:eff}.

\begin{table*}
 \caption{
 The reconstruction efficiencies for $D^+\to \bar K^0e^+\nu_e$ and
 $D^+\to \pi^0e^+\nu_e$ determined from MC simulation.
 The efficiencies include the branching fractions for $\bar K^0$ and $\pi^0$. The uncertainties are statistical only.
 }
 \label{tab:eff}
 \begin{ruledtabular}
 \begin{tabular}{lcccc}
  Tag mode
  & $\varepsilon_{\rm tag,SL}(D^+\to \bar K^0e^+\nu_e)$ (\%)
  & $\varepsilon(D^+\to \bar K^0e^+\nu_e)$ (\%)
  & $\varepsilon_{\rm tag,SL}(D^+\to \pi^0e^+\nu_e)$ (\%)
  & $\varepsilon(D^+\to \pi^0e^+\nu_e)$ (\%) \\
  \hline
$D^-\to K^+\pi^-\pi^-$               & $~9.21\pm0.02$ & $17.77\pm0.04$   & $28.44\pm0.06$ & $54.88\pm0.13$ \\
$D^-\to K^0_S\pi^-$                  & $10.14\pm0.05$ & $18.05\pm0.11$   & $31.15\pm0.15$ & $55.43\pm0.34$ \\
$D^-\to K^0_S K^-$                   & $~9.30\pm0.08$ & $17.84\pm0.22$   & $28.68\pm0.23$ & $55.02\pm0.67$ \\
$D^-\to K^+K^-\pi^-$                 & $~7.39\pm0.06$ & $17.92\pm0.18$   & $22.53\pm0.16$ & $54.66\pm0.53$ \\
$D^-\to K^+\pi^-\pi^-\pi^0$          & $~4.98\pm0.02$ & $18.25\pm0.09$   & $15.49\pm0.06$ & $56.72\pm0.29$ \\
$D^-\to \pi^+\pi^-\pi^-$             & $10.44\pm0.11$ & $18.34\pm0.30$   & $32.93\pm0.33$ & $57.82\pm0.94$ \\
$D^-\to K^0_S\pi^-\pi^0$             & $~5.67\pm0.01$ & $18.11\pm0.08$   & $17.83\pm0.04$ & $56.92\pm0.25$ \\
$D^-\to K^+\pi^-\pi^-\pi^-\pi^+$     & $~3.50\pm0.04$ & $15.88\pm0.25$   & $11.74\pm0.14$ & $53.20\pm0.81$ \\
$D^-\to K^0_S\pi^-\pi^-\pi^+$        & $~5.55\pm0.02$ & $16.84\pm0.14$   & $18.12\pm0.06$ & $54.97\pm0.45$ \\
\hline
  Average                     &                & $17.83\pm0.03$   &                & $55.52\pm0.10$ \\
 \end{tabular}
 \end{ruledtabular}
\end{table*}

Using the control samples selected from Bhabha scattering and $D\bar D$ events,
we find that there are small discrepancies between data and MC simulation in the
positron tracking efficiency, positron identification efficiency,
$K_S^0$ and $\pi^0$ reconstruction efficiencies.
We correct for these differences by multiplying the raw efficiencies
$\varepsilon(D^+\to\bar K^0e^+\nu_e)$ and $\varepsilon(D^+\to\pi^0e^+\nu_e)$
determined in MC simulation by factors of 0.9957 and 0.9910, respectively.
The corrected efficiencies are found to be
$\epsilon^\prime(D^+\to\bar K^0e^+\nu_e)=(17.75\pm0.03)\%$ and
$\epsilon^\prime(D^+\to\pi^0e^+\nu_e)=(55.02\pm0.10)\%$,
where the uncertainties are only statistical.

Inserting the corresponding numbers into Eq.~(\ref{eq:BF})
yields the absolute decay branching fractions
\begin{linenomath*}
\begin{equation}
 \mathcal B(D^+\to\bar K^0e^+\nu_e) = (8.60\pm 0.06\pm 0.15)\times10^{-2}
\label{eq:BF_DptoK0enu}
\end{equation}
\end{linenomath*}
and
\begin{linenomath*}
\begin{equation}
 \mathcal B(D^+\to\pi^0e^+\nu_e)=(3.63\pm 0.08\pm 0.05)\times10^{-3},
\label{eq:BF_pi0enu}
\end{equation}
\end{linenomath*}
where the first uncertainties are statistical and the second systematic.

\subsection{Systematic uncertainties}

The systematic uncertainties in the measured branching fractions of
$D^+\to\bar K^0e^+\nu_e$ and $D^+\to\pi^0e^+\nu_e$ decays
include the following contributions.

\textit{Number of $D^-$ tags.}
The systematic uncertainty of the number of $D^-$ tags is 0.5\%~\cite{BESIII_Dptomunu}.

\textit{$e^+$ tracking efficiency.}
Using the positron samples selected from radiative Bhabha scattering events,
the $e^+$ tracking efficiencies are measured in data and MC simulation.
Considering both the polar angle and momentum distributions of the positrons
in the semileptonc decays,
a correction factor of $1.0021\pm0.0019$ ($1.0011\pm0.0015$)
is determined for the $e^+$ tracking efficiency in the branching fraction measurement of $D^+\to\bar K^0e^+\nu_e$ ($D^+\to\pi^0e^+\nu_e$) decay.
This correction is applied and an uncertainty of 0.19\% (0.15\%) is
used as the corresponding systematic uncertainty.

\textit{$e^+$ identification efficiency.}
Using the positron samples selected from radiative Bhabha scattering events,
we measure the $e^+$ identification efficiencies in data and MC simulation.
Taking both the polar angle and momentum distributions of the positrons
in the semileptonic decays into account,
a correction factor of $0.9993\pm0.0016$ ($0.9984\pm0.0014$)
is determined for the $e^+$ identification efficiency
in the measurement of $\mathcal B(D^+\to\bar K^0e^+\nu_e)$
($\mathcal B(D^+\to\pi^0e^+\nu_e)$).
This correction is applied, and an amount of 0.16\% (0.14\%) is
assigned as the corresponding systematic uncertainty.

\textit{$K_S^0$ and $\pi^0$ reconstruction efficiency.}
The momentum-dependent efficiencies for $K^0_S$ ($\pi^0$) reconstruction
in data and in MC simulation are measured with $D \bar D$ events.
Weighting these efficiencies according to the $K^0_S$ ($\pi^0$) momentum distribution
in the semileptonic decay
leads to a difference of $(-0.57\pm1.62)\%$ ($(-0.85\pm1.00)\%$) between
the $K_S^0$ ($\pi^0$) reconstruction efficiencies in data and MC simulation.
Since we correct for the systematic shift,
the uncertainty of the correction factor, $1.62\%$ ($1.00\%$), is
taken as the corresponding systematic uncertainty in the measured
branching fraction of $D^+\to\bar K^0e^+\nu_e$ ($D^+\to\pi^0e^+\nu_e$).

\textit{Requirement on $E_{\gamma, \rm max}$.}
By comparing doubly tagged $D\bar D$ hadronic decay events in the data and MC
simulation, the systematic uncertainty due to this source is estimated to be 0.1\%.

\textit{Fit to the $U_{\rm miss}$ distribution.}
To estimate the uncertainties
due to the fits to the $U_{\rm miss}$ distributions,
we refit the $U_{\rm miss}$ distributions by varying the bin size and the tail parameters
(which are used to describe the signal shapes and are determined from MC simulation)
to obtain the number of signal events from $D^+$ semileptonic decays.
We then combine the changes in the yields in quadrature to obtain
the systematic uncertainty
(0.12\% for $D^+\to\bar K^0e^+\nu_e$, 0.52\% for $D^+\to\pi^0e^+\nu_e$).
Since the background function
is formed from many background modes with fixed relative
normalizations, we also vary the relative contributions of
several of the largest background modes based on the
uncertainties in their branching fractions
(0.12\% for $D^+\to\bar K^0e^+\nu_e$, 0.01\% for $D^+\to\pi^0e^+\nu_e$).
In addition, we convolute the background shapes formed from MC simulation
with the same Gaussian function in the fits
(0.02\% for $D^+\to\bar K^0e^+\nu_e$, 0.30\% for $D^+\to\pi^0e^+\nu_e$).
Finally we assign the relative uncertainties to be $0.2\%$ and $0.6\%$ for
$D^+ \to \bar K^0 e^+ \nu_e$ and $D^+ \to \pi^0 e^+ \nu_e$, respectively.

\textit{Form factor.}
In order to estimate the systematic uncertainty associated with
the form factor used to generate signal events in the MC simulation,
we re-weight the signal MC events so that the $q^2$ spectra
agree with the measured spectra. We then remeasure the branching fraction
(partial decay rates in different $q^2$ bins)
with the newly weighted efficiency (efficiency matrix).
The maximum relative change of the branching fraction
(partial decay rates in different $q^2$ bins) is $0.2\%$
and is assigned as the systematic uncertainty.%

\textit{FSR recovery.}
The differences between the results with FSR recovery and the ones without FSR
recovery are assigned as the systematic uncertainties due to FSR recovery.
We find the differences are $0.1\%$ and $0.5\%$ for $D^+\to \bar K^0e^+\nu_e$
and $D^+\to \pi^0e^+\nu_e$, respectively.

\textit{MC statistics.}
The uncertainties in the measured branching fractions due to the MC statistics are
the statistical fluctuation of the MC samples, which are $0.2\%$ for both of
$D^+ \rightarrow \bar K^0 e^+\nu_e$ and $D^+ \rightarrow \pi^0 e^+\nu_e$ semileptonic decays.

\textit{$K_S^0$ and $\pi^0$ decay branching fractions.}
We include an uncertainty of 0.07\% (0.03\%) on the branching fraction measurement
of $D^+\to\bar K^0 e^+\nu_e$ ($D^+\to\pi^0 e^+\nu_e$)
to account for the uncertainty of the branching fraction of $K_S^0 \to \pi^+ \pi^-$
($\pi^0 \to \gamma\gamma$) decay~\cite{pdg2016}.

\begin{table}
\caption{
Summary of the systematic uncertainties considered
in the measurements of the branching fractions of
$D^+ \rightarrow \bar K^0 e^+\nu_e$ and $D^+ \rightarrow \pi^0 e^+\nu_e$ decays.}
\label{tab:syst}
\begin{ruledtabular}
\begin{tabular}{lcc}
& \multicolumn{2}{c} {Systematic uncertainty (\%)} \\
Source  & $D^+\to \bar K^0e^+\nu_e$  & $D^+\to\pi^0e^+\nu_e$  \\
\hline
 Number of $D^-$ tags                  & 0.5       &    0.5      \\
 Tracking for $e^+$                    & 0.19      &    0.15     \\
 PID for $e^+$                         & 0.16      &    0.14     \\
 $K_S^0$ reconstruction                & 1.62      &    $\cdots$ \\
 $\pi^0$ reconstruction                & $\cdots$  &    1.00     \\
 Requirement on $E_{\gamma,\rm max}$   & 0.1       &    0.1      \\
 Fit to $U_{\rm miss}$ distribution    & 0.2       &    0.6      \\
 Form factor                           & 0.2       &    0.2      \\
 FSR recovery                          & 0.1       &    0.5      \\
 MC statistics                         & 0.2       &    0.2      \\
 $K^0_S/\pi^0$ branching fraction      & 0.07      &    0.03     \\
\hline
 Total                                & 1.76       &    1.41     \\
\end{tabular}
\end{ruledtabular}
\end{table}

Table~\ref{tab:syst} summarizes the systematic uncertainties
in the measurement of the branching fractions.
Adding all systematic uncertainties in quadrature yields the total systematic uncertainties
of $1.76\%$ and $1.41\%$ for $D^+\to\bar K^0 e^+\nu_e$ and $D^+\to\pi^0 e^+\nu_e$, respectively.

\subsection{Comparison}

\begin{figure*}
\centerline{
\includegraphics[width=0.48\textwidth]{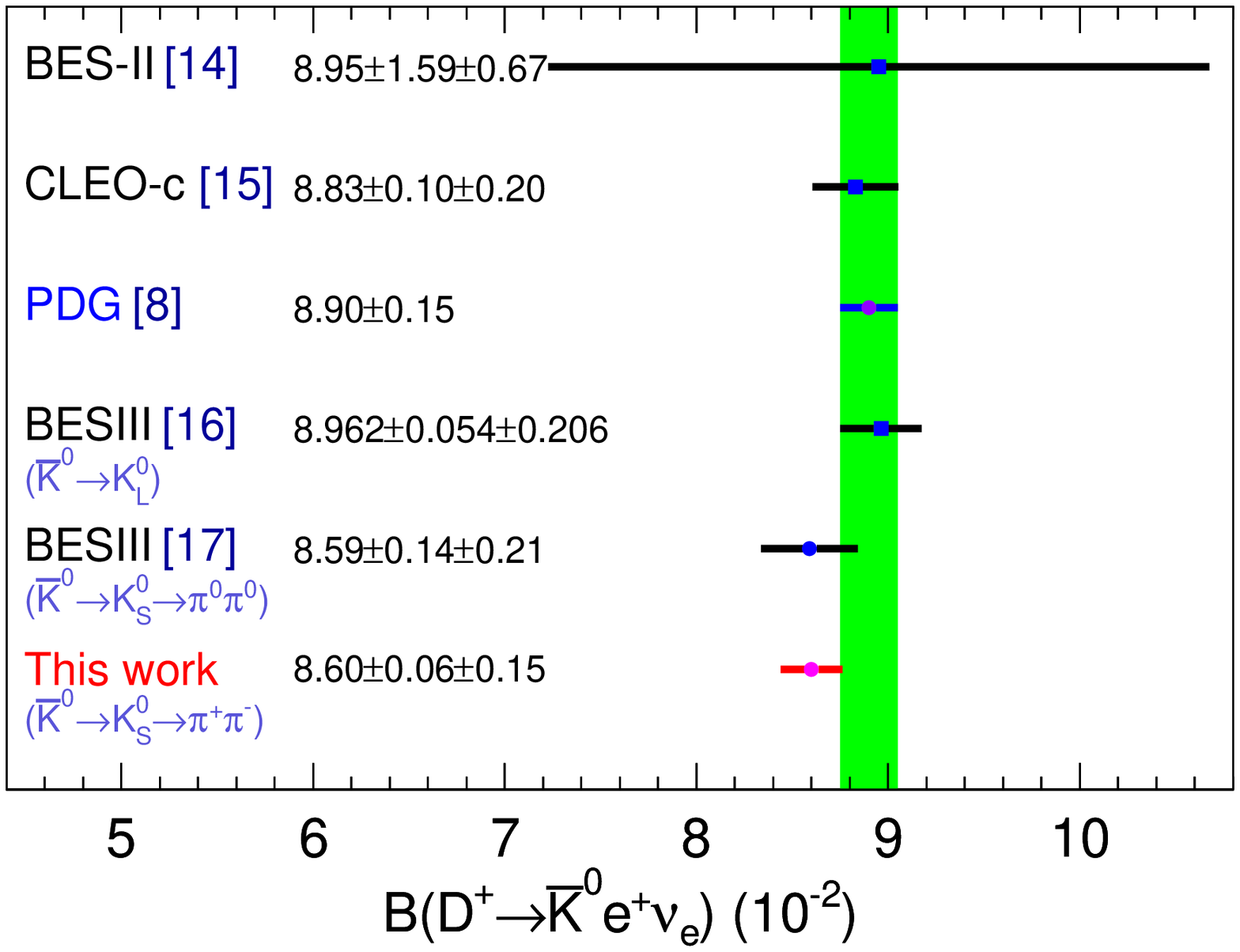}
\includegraphics[width=0.48\textwidth]{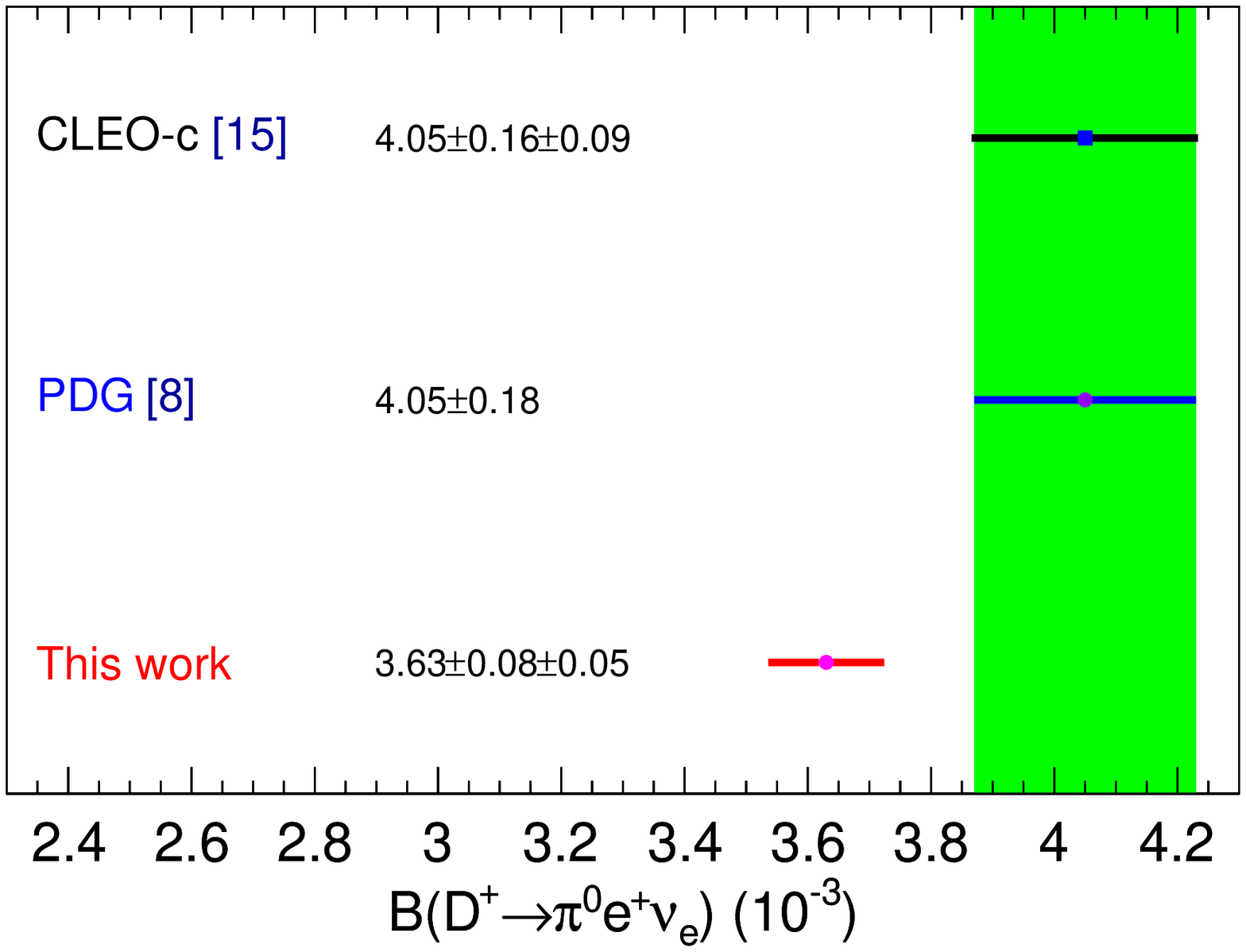}
}
\caption{
Comparison of the branching fraction measurements for
$D^+\to\bar K^0e^+\nu_e$ (left) and $D^+\to\pi^0e^+\nu_e$ (right).
The green bands correspond to the $1\sigma$ limits of the world averages.
}
\label{fig:Cmp_BF}
\end{figure*}

The comparisons of our measured branching fractions
for $D^+\to\bar K^0e^+\nu_e$ and
$D^+\to\pi^0e^+\nu_e$ decays with those
previously measured at the
BES-II~\cite{BESII_DptoK0enu},
CLEO-c~\cite{CLEOc_DtoPenu_818}
and BESIII~\cite{BESIII_DptoKLenu,BESIII_DptoKSenu} experiments
as well as the PDG values~\cite{pdg2016}
are shown in Fig.~\ref{fig:Cmp_BF}.
Our measured branching fractions are in agreement
with the other experimental measurements, but are more precise.
For $D^+\to\pi^0e^+\nu_e$, our result is lower than the only other existing measurement by CLEO-c~\cite{CLEOc_DtoPenu_818} by $2.0\sigma$.

Using our previous measurements of $\mathcal B(D^0\to K^-e^+\nu_e)$ and
$\mathcal B(D^0\to\pi^-e^+\nu_e)$~\cite{BESIII_D0toPenu},
the results obtained in this analysis,
and the lifetimes of $D^0$ and $D^+$ mesons~\cite{pdg2016},
we obtain the ratios
\begin{equation}\label{eq:I_K}
    I_K \equiv \frac{\Gamma(D^0\to K^-e^+\nu_e)}{\Gamma(D^+\to\bar K^0e^+\nu_e)}
    =1.03\pm0.01\pm0.02
\end{equation}
and
\begin{equation}\label{eq:I_pi}
    I_\pi \equiv \frac{\Gamma(D^0\to\pi^-e^+\nu_e)}{2\Gamma(D^+\to\pi^0e^+\nu_e)}
    =1.03\pm0.03\pm0.02,
\end{equation}
which are consistent with isospin symmetry.

\section{Partial decay rate measurements}
\label{sec:DR}

\subsection{Determinations of partial decay rates}

To study the differential decay rates,
we divide the semileptonic candidates satisfying the selection criteria described in
Sec.~\ref{sec:sel} into bins of $q^2$. Nine (seven) bins are used for
$D^+\to\bar K^0e^+\nu_e$ ($D^+\to\pi^0e^+\nu_e$). The range of each bin is given in
Table~\ref{tab:Nobs_i}.
The squared four momentum transfer $q^{2}$ is determined for each semileptonic candidate by
$q^2 = (E_{e^+}+E_{\nu_e})^2/c^4 - (\vec p_{e^+} + \vec p_{\nu_e})^2/c^2$,
where the energy and momentum of the missing neutrino are taken to be
$E_{\nu_e} = E_{\rm miss}$ and
$\vec p_{\nu_e} = E_{\rm miss}\hat p_{\rm miss}/c$, respectively.
For each $q^2$ bin, we perform a maximum likelihood fit to the corresponding
$U_{\rm miss}$ distribution following the same procedure described in Sec.~\ref{sec:sel:sl} and obtain the signal yields as shown in Table~\ref{tab:Nobs_i}.

\begin{table*}
\centering
\caption{
Summary of the range of each $q^2$ bin,
the number of the observed signal events for $D^+\to \bar K^0e^+\nu_e$ and $D^+\to \pi^0e^+\nu_e$ in data.
}
\label{tab:Nobs_i}
\begin{ruledtabular}
\begin{tabular}{lccccccccc}
\multicolumn{10}{c}{$D^+\to\bar K^0e^+\nu_e$} \ST \\
Bin No. &  1 &  2 &  3 &  4 &  5 &  6 &  7 &  8 &  9 \\
$q^2$ (GeV$^2/c^4)$ & $[0.0, 0.2)$ & $[0.2, 0.4)$ & $[0.4, 0.6)$ & $[0.6, 0.8)$ & $[0.8, 1.0)$ & $[1.0, 1.2)$ & $[1.2, 1.4)$ & $[1.4, 1.6)$ & $[1.6, q^2_{\rm max})$ \\
$N_{\rm obs}$ & $5842\pm81$ & $4935\pm73$ & $4180\pm67$ & $3515\pm62$ & $2818\pm55$ & $2120\pm48$ & $1460\pm40$ & $ 860\pm31$ & $ 302\pm19$ \\
\hline
\multicolumn{10}{c}{$D^+\to\pi^0e^+\nu_e$} \ST \\
Bin No. &  1 &  2 &  3 &  4 &  5 &  6 &  7 \\
$q^2$ (GeV$^2/c^4)$ & $[0.0, 0.3)$& $[0.3, 0.6)$& $[0.6, 0.9)$& $[0.9, 1.2)$& $[1.2, 1.5)$& $[1.5, 2.0)$& $[2.0, q^2_{\rm max})$ \\
$N_{\rm obs}$  & $658 \pm 29$ & $562 \pm 27$ & $467 \pm 25$ & $448 \pm 24$ & $401 \pm 24$ & $470 \pm 26$ & $404 \pm 30$\\
\end{tabular}
\end{ruledtabular}
\end{table*}

To account for detection efficiency and detector resolution,
the number of events $N^i_{\rm obs}$ observed in the $i$th $q^2$ bin is
extracted from the relation
\begin{linenomath*}
\begin{equation}\label{eq:Nobs_i}
N^{i}_{\rm obs}=\sum_{j=1}^{N_{\rm bins}}\varepsilon_{ij}N^{j}_{\rm prd},
\end{equation}
\end{linenomath*}
where
$N_{\rm bins}$ is the number of $q^2$ bins,
$N_{\rm prd}^j$ is the number of semileptonic decay events produced in the tagged
$D^-$ sample with the $q^2$ filled in the $j$th bin, and
$\varepsilon_{ij}$ is the overall efficiency matrix that describes the efficiency
and smearing across $q^{2}$ bins.
The efficiency matrix element $\varepsilon_{ij}$ is obtained by
\begin{linenomath*}
\begin{equation}
 \varepsilon_{ij} = \frac{n^{\rm rec}_{ij}}{n^{\rm gen}_j}
\frac{1}{\varepsilon_{\rm tag}}
f_{ij},
\end{equation}
\end{linenomath*}
where $n^{\rm rec}_{ij}$ is the number of the signal MC events generated in the $j$th $q^2$ bin
and reconstructed in the $i$th $q^2$ bin,
$n^{\rm gen}_j$ is the total number of the signal MC events which are generated in the $j$th $q^2$ bin,
and $f_{ij}$ is the matrix
to correct for data-MC differences in the efficiencies for $e^+$ tracking, $e^+$ identification,
and $\bar K^0$ ($\pi^0$) reconstruction.
Table~\ref{tab:eff_mat} presents the average overall efficiency matrices for $D^+ \to \bar{K^0}e^{+}\nu_{e}$ and $D^+ \to \pi^0 e^+ \nu_e$ decays.
To produce this average overall efficiency matrix, we combine the efficiency matrices for
each tag mode weighted by its yield shown in Table~\ref{tab:Dtag}.
The diagonal elements of the matrix give the overall efficiencies for $D^+\to Pe^+\nu_e$ decays
to be reconstructed in the correct $q^2$ bins in the recoil of the single $D^-$ tags,
while the neighboring off-diagonal elements of the matrix give the overall efficiencies
for cross feed between different $q^2$ bins.

\begin{table*}
\centering
 \caption{
 Efficiency matrices $\varepsilon_{ij}$ given in percent for $D^+ \to \bar{K^0}e^{+}\nu_{e}$ and $D^+ \to \pi^0 e^+ \nu_e$ decays. The column gives the true $q^{2}$ bin $j$, while the row gives the reconstructed $q^{2}$ bin $i$.
 The statistical uncertainties in the least significant digits are given in the parentheses.
 }
 \label{tab:eff_mat}
\begin{ruledtabular}
 \begin{tabular}{lccccccccc}
  \multicolumn{10}{c}{$D^+ \to \bar K^0 e^+ \nu_e$} \ST \\
  Rec. $q^{2}$ & \multicolumn{2}{l}{True $q^{2}$ (GeV$^{2}/c^{4}$)} \\
  (GeV$^{2}/c^{4}$) & $[0.0,0.2)$ & $[0.2,0.4)$ & $[0.4,0.6)$ & $[0.6,0.8)$ & $[0.8,1.0)$ & $[1.0,1.2)$ & $[1.2,1.4)$ & $[1.4,1.6)$ & $[1.6,q^2_{\rm max})$ \\
  $[0.0,0.2)$           & $18.53(6)$  & $0.95(1)$  & $0.07(0)$  & $0.00(0)$  & $0.00(0)$  & $0.00(0)$  & $0.00(0)$  & $0.00(0)$  & $0.00(0)$ \\
  $[0.2,0.4)$           & $0.37(1)$  & $16.86(6)$  & $1.03(2)$  & $0.05(0)$  & $0.00(0)$  & $0.00(0)$  & $0.00(0)$  & $0.00(0)$  & $0.00(0)$ \\
  $[0.4,0.6)$           & $0.00(0)$  & $0.40(1)$  & $16.03(6)$  & $1.03(2)$  & $0.03(0)$  & $0.00(0)$  & $0.00(0)$  & $0.00(0)$  & $0.00(0)$ \\
  $[0.6,0.8)$           & $0.00(0)$  & $0.00(0)$  & $0.46(1)$  & $15.72(6)$  & $0.95(2)$  & $0.02(0)$  & $0.00(0)$  & $0.00(0)$  & $0.00(0)$ \\
  $[0.8,1.0)$           & $0.00(0)$  & $0.00(0)$  & $0.01(0)$  & $0.44(1)$  & $15.78(7)$  & $0.93(2)$  & $0.01(0)$  & $0.00(0)$  & $0.00(0)$ \\
  $[1.0,1.2)$           & $0.00(0)$  & $0.00(0)$  & $0.00(0)$  & $0.01(0)$  & $0.46(1)$  & $15.76(8)$  & $0.80(2)$  & $0.01(0)$  & $0.00(0)$ \\
  $[1.2,1.4)$           & $0.00(0)$  & $0.00(0)$  & $0.00(0)$  & $0.00(0)$  & $0.00(0)$  & $0.42(1)$  & $15.58(9)$  & $0.74(3)$  & $0.00(0)$ \\
  $[1.4,1.6)$           & $0.00(0)$  & $0.00(0)$  & $0.00(0)$  & $0.00(0)$  & $0.00(0)$  & $0.00(0)$  & $0.38(2)$  & $15.45(12)$  & $0.78(5)$ \\  $[1.6,q^2_{\rm max})$ & $0.00(0)$  & $0.00(0)$  & $0.00(0)$  & $0.00(0)$  & $0.00(0)$  & $0.00(0)$  & $0.00(0)$  & $0.28(2)$  & $15.98(19)$ \\
  \hline
  \multicolumn{10}{c}{$D^+ \to \pi^0 e^+ \nu_e$} \ST \\
  Rec. $q^{2}$ & \multicolumn{2}{l}{True $q^{2}$ (GeV$^{2}/c^{4}$)} \\
  (GeV$^{2}/c^{4}$) & $[0.0,0.3)$ & $[0.3,0.6)$ & $[0.6,0.9)$ & $[0.9,1.2)$ & $[1.2,1.5)$ & $[1.5,2.0)$ & $[2.0,q^2_{\rm max})$ \\
 $[0.0,0.3)$             & $53.84(15)$  & $2.27(3)$  & $0.17(1)$  & $0.01(0)$  & $0.00(0)$  & $0.00(0)$  & $0.00(0)$ \\
 $[0.3,0.6)$             & $4.00(5)$  & $48.24(15)$  & $2.31(4)$  & $0.14(1)$  & $0.00(0)$  & $0.00(0)$  & $0.01(0)$ \\
 $[0.6,0.9)$             & $0.14(1)$  & $5.66(6)$  & $46.15(15)$  & $2.34(4)$  & $0.10(1)$  & $0.00(0)$  & $0.00(0)$ \\
 $[0.9,1.2)$             & $0.04(0)$  & $0.22(1)$  & $6.24(6)$  & $44.51(16)$  & $2.16(4)$  & $0.05(0)$  & $0.00(0)$ \\
 $[1.2,1.5)$             & $0.04(0)$  & $0.08(1)$  & $0.31(1)$  & $6.33(7)$  & $43.33(17)$  & $1.36(3)$  & $0.02(0)$ \\
 $[1.5,2.0)$             & $0.03(0)$  & $0.08(1)$  & $0.22(1)$  & $0.58(2)$  & $6.52(8)$  & $45.48(16)$  & $1.12(3)$ \\
 $[2.0,q^2_{\rm max})$  & $0.13(1)$  & $0.21(1)$  & $0.34(1)$  & $0.68(2)$  & $1.30(3)$  & $5.52(6)$  & $50.46(19)$ \\
 \end{tabular}
\end{ruledtabular}
\end{table*}

The partial decay width in the $i$th bin is  obtained by
inverting the matrix Eq.~(\ref{eq:Nobs_i}),
\begin{linenomath*}
\begin{equation}
\Delta\Gamma_{i}=\frac{N_{\rm prd}^{i}}{\tau_{D^+} N_{\rm tag}}
=\frac{1}{\tau_{D^+} N_{\rm tag}}\sum_{j}^{N_{\rm bins}}(\varepsilon^{-1})_{ij}N_{\rm obs}^{j},
\label{eq:DR_i}
\end{equation}
\end{linenomath*}
where $\tau_{D^+}$ is the lifetime of the $D^+$ meson~\cite{pdg2016}.
The $q^2$-dependent partial widths for $D^+\to\bar K^0e^+\nu_e$ and $D^+\to\pi^0e^+\nu_e$
are summarized in Table~\ref{tab:DR}.
Also shown in Table~\ref{tab:DR} are the statistical uncertainties and
the associated correlation matrices.

\begin{table*}
 \caption{Summary of the measured partial decay rates, relative statistical uncertainties,
 systematic uncertainties and corresponding correlation matrices
 for $D^+\to \bar K^0e^+\nu_e$ and $D^+\to \pi^0e^+\nu_e$.
 }
 \label{tab:DR}
 \centering
 \begin{ruledtabular}
 \begin{tabular}{lrrrrrrrrr}
  \multicolumn{10}{c}{$D^+\to \bar K^0 e^+\nu_e$} \\
   $q^2$ bin No. &  1~~~ &  2~~~ &  3~~~ &  4~~~ &  5~~~ &  6~~~ &  7~~~ &  8~~~ &  9~~~ \\
   $\Delta\Gamma$ (ns$^{-1}$) & $16.97$  & $15.29$  & $13.57$  & $11.65$  & $ 9.33$  & $ 7.06$  & $ 4.96$  & $ 2.97$  & $ 1.01$ \\
   stat. uncert. (\%) & $1.45$ & $1.61$ & $1.75$ & $1.91$ & $2.12$ & $2.44$ & $2.92$ & $3.77$ & $6.56$ \\
   stat. correl.
      & $  1.000$  \\
      & $ -0.073$ & $  1.000$  \\
      & $  0.001$ & $ -0.084$ & $  1.000$  \\
      & $  0.000$ & $  0.003$ & $ -0.091$ & $  1.000$   \\
      & $  0.000$ & $  0.000$ & $  0.004$ & $ -0.085$ & $  1.000$  \\
      & $  0.000$ & $  0.000$ & $  0.000$ & $  0.004$ & $ -0.085$ & $  1.000$  \\
      & $  0.000$ & $  0.000$ & $  0.000$ & $  0.000$ & $  0.004$ & $ -0.075$ & $  1.000$  \\
      & $  0.000$ & $  0.000$ & $  0.000$ & $  0.000$ & $  0.000$ & $  0.004$ & $ -0.069$ & $  1.000$  \\
      & $  0.000$ & $  0.000$ & $  0.000$ & $  0.000$ & $  0.000$ & $  0.000$ & $  0.003$ & $ -0.059$ & $  1.000$  \\
   syst. uncert. (\%) & $3.24$ & $3.10$ & $2.95$ & $2.88$ & $3.02$ & $3.05$ & $2.85$ & $2.54$ & $2.93$ \\
   syst. correl.
 & $  1.000$ \\
 & $  0.981$ & $  1.000$ \\
 & $  0.979$ & $  0.976$ & $  1.000$ \\
 & $  0.979$ & $  0.977$ & $  0.973$ & $  1.000$ \\
 & $  0.978$ & $  0.976$ & $  0.973$ & $  0.970$ & $  1.000$ \\
 & $  0.974$ & $  0.972$ & $  0.970$ & $  0.970$ & $  0.965$ & $  1.000$ \\
 & $  0.966$ & $  0.964$ & $  0.963$ & $  0.962$ & $  0.960$ & $  0.954$ & $  1.000$ \\
 & $  0.932$ & $  0.930$ & $  0.929$ & $  0.929$ & $  0.926$ & $  0.923$ & $  0.911$ & $  1.000$ \\
 & $  0.891$ & $  0.889$ & $  0.886$ & $  0.888$ & $  0.886$ & $  0.883$ & $  0.875$ & $  0.840$ & $  1.000$ \\
  \hline
  \multicolumn{10}{c}{$D^+\to \pi^0 e^+\nu_e$} \\
  $q^2$ bin No. &  1~~~ &  2~~~ &  3~~~ &  4~~~ &  5~~~ &  6~~~ &  7~~~   \\
  $\Delta\Gamma$ (ns$^{-1}$)  & $0.664$  & $0.578$  & $0.474$  & $0.477$  & $0.432$  & $0.503$ & $0.372$  \\
  stat. uncert. (\%) & $4.55$ & $5.53$ & $6.60$ & $6.48$ & $7.28$ & $6.52$ & $8.97$ \\
  stat. correl.
      & $  1.000$    \\
      & $ -0.122$ & $  1.000$    \\
      & $  0.011$ & $ -0.171$ & $  1.000$    \\
      & $ -0.002$ & $  0.019$ & $ -0.190$ & $  1.000$    \\
      & $  0.000$ & $ -0.003$ & $  0.021$ & $ -0.190$ & $  1.000$    \\
      & $  0.000$ & $ -0.001$ & $ -0.005$ & $  0.016$ & $ -0.167$ & $  1.000$    \\
      & $ -0.002$ & $ -0.003$ & $ -0.003$ & $ -0.008$ & $ -0.004$ & $ -0.128$ & $  1.000$   \\
  syst. uncert. (\%) & $1.53$ & $1.52$ & $1.51$ & $1.61$ & $1.88$ & $1.92$ & $1.73$ \\
  syst. correl.
 & $  1.000$ \\
 & $  0.739$ & $  1.000$ \\
 & $  0.742$ & $  0.664$ & $  1.000$ \\
 & $  0.758$ & $  0.737$ & $  0.650$ & $  1.000$ \\
 & $  0.772$ & $  0.740$ & $  0.712$ & $  0.698$ & $  1.000$ \\
 & $  0.781$ & $  0.749$ & $  0.711$ & $  0.760$ & $  0.772$ & $  1.000$ \\
 & $  0.760$ & $  0.730$ & $  0.697$ & $  0.727$ & $  0.756$ & $  0.740$ & $  1.000$ \\
 \end{tabular}
 \end{ruledtabular}
\end{table*}

\subsection{Systematic covariance matrices}

For each source of systematic uncertainty in the measurements of partial decay rates,
we construct an $N_{\rm bins} \times N_{\rm bins}$ systematic covariance matrix.
A brief description of each contribution follows.

  \textit{$D^+$ lifetime.}
  The systematic uncertainty associated with the lifetime of the $D^+$ meson
  (0.7\%)~\cite{pdg2016}
  is fully correlated across $q^2$ bins.

  \textit{Number of $D^-$ tags.}
  The systematic uncertainty from the number of the single $D^-$ tags (0.5\%)
  is fully correlated between $q^2$ bins.

  \textit{$e^+$, $K_S^0$, and $\pi^0$ reconstruction.}
  The covariance matrices for the systematic uncertainties associated with the
  $e^+$ tracking, $e^+$ identification, $K_S^0$, and $\pi^0$ reconstruction
  efficiencies are obtained in the following way.
  We first vary the corresponding correction factors according to
  their uncertainties, then remeasure
  the partial decay rates using the efficiency matrices determined from the re-corrected signal MC events.
  The covariance matrix due to this source is assigned via
  $C_{ij}=\delta(\Delta\Gamma_i)\delta(\Delta\Gamma_j)$,
  where $\delta(\Delta\Gamma_i)$ denotes the change in the partial decay rate measurement in the $i$th $q^2$ bin.

  \textit{Requirement on $E_{\gamma,\rm max}$.}
  We take the systematic uncertainty of $0.1\%$ due to the $E_{\gamma,\rm max}$
  requirement on the selected events in each $q^2$ bin,
  and assume that this uncertainty is fully correlated between $q^2$ bins.

  \textit{Fit to the $U_{\rm miss}$ distribution.}
  The technique of fitting the $U_{\rm miss}$ distributions affects the number of signal events observed in the $q^2$ bins.
  The covariance matrix due to the $U_{\rm miss}$ fits is determined by
  \begin{linenomath*}
  \begin{equation}
   C_{ij}= \left(\frac{1}{\tau_{D^+} N_{\rm tag}}\right)^{2}\sum_{\alpha} \epsilon^{-1}_{i \alpha}\epsilon^{-1}_{j \alpha} [\delta(N_{\rm obs}^\alpha)]^2,
  \end{equation}
  \end{linenomath*}
  where $\delta(N_{\rm obs}^\alpha)$ is the systematic uncertainty of
  $N_{\rm obs}^\alpha$ associated with the fit to the corresponding $U_{\rm miss}$ distribution.

  \textit{Form factor.}
  To estimate the systematic uncertainty associated with
  the form factor model used to generate signal events in the MC simulation,
  we re-weight the signal MC events so that the $q^2$ spectra
  agree with the measured spectra.
  We then re-calculate the partial decay rates in different $q^2$ bins
  with the new efficiency matrices which are determined using the weighted
  MC events. The covariance matrix due to this source is assigned via
  $C_{ij}=\delta(\Delta\Gamma_i)\delta(\Delta\Gamma_j)$,
  where $\delta(\Delta\Gamma_i)$ denotes the change of the partial width measurement in the $i$th $q^2$ bin.

  \textit{FSR recovery.}
  To estimate the systematic covariance matrix associated with the FSR recovery of the positron momentum,
  we remeasure the partial decay rates without the FSR recovery.
  The covariance matrix due to this source is assigned via
  $C_{ij}=\delta(\Delta\Gamma_i)\delta(\Delta\Gamma_j)$,
  where $\delta(\Delta\Gamma_i)$ denotes the change of the partial decay rate measurement in the $i$th $q^2$ bin.

  \textit{MC statistics.}
  The systematic uncertainties due to the limited size of the MC samples
  used to determine the efficiency matrices are translated to the covariance via
  \begin{linenomath*}
  \begin{equation}
    C_{ij} = \left(\frac{1}{\tau_{D^+} N_{\rm tag}}\right)^{2} \sum_{\alpha \beta}
    (N_{\rm obs}^{\alpha}N_{\rm obs}^{\beta} {\rm cov}[\epsilon^{-1}_{i \alpha},\epsilon^{-1}_{j \beta}] ),
  \end{equation}
  \end{linenomath*}
  where the covariance of the inverse efficiency matrix elements are given
  by~\cite{cov_inverse_matrix}
  \begin{linenomath*}
   \begin{equation}
    {\rm cov}[\epsilon^{-1}_{\alpha\beta},\epsilon^{-1}_{ab}]
    = \sum_{ij}(\epsilon^{-1}_{\alpha i}\epsilon^{-1}_{ai})
    [\sigma^2(\epsilon_{ij})]^2
       (\epsilon^{-1}_{j \beta}\epsilon^{-1}_{jb}).
   \end{equation}
  \end{linenomath*}

  \textit{$K_S^0$ and $\pi^0$ decay branching fractions.}
  The systematic uncertainties due to the branching fractions of
  $K_S^0 \to \pi^+\pi^-$ (0.07\%) and
  $\pi^0 \to \gamma\gamma$ (0.03\%) are fully correlated between $q^2$ bins.

The total systematic covariance matrix is obtained by summing all these matrices.
Table~\ref{tab:DR} summarizes
the relative size of systematic uncertainties and the corresponding
correlations in the measurements for the partial decay rates of the
$D^+\to\bar K^0e^+\nu_e$ and $D^+\to\pi^0e^+\nu_e$ semileptonic decays.

\section{Form Factors}
\label{sec:FF}

To determine the product $f_+(0)|V_{cs(d)}|$ and
other form factor parameters,
we fit the measured partial decay rates using Eq.~(\ref{eq:dG_dq2})
with the parameterization of the form factor $f_+(q^2)$.
In this analysis, we use several forms of the form factor
parameterizations which are reviewed in Sec.~\ref{sec:form_facor}.

\subsection{Form factor parameterizations}
\label{sec:form_facor}

In general, the \textit{single pole model} is the simplest approach to describe the $q^2$ dependence of the form factor. The single pole model is expressed as
\begin{equation}\label{eq:ff_pole}
    f_+(q^2) = \frac{f_+(0)}{1-q^2/m_{\rm pole}^2},
\end{equation}
where $f_+(0)$ is the value of the form factor at $q^2=0$, and $m_{\rm pole}$ is the pole
mass, which is often treated as a free parameter to improve fit quality.

The \textit{modified pole model}~\cite{BK} is also widely used in Lattice QCD (LQCD) calculations
and experimental studies of these decays.
In this parameterization, the form factor
of the semileptonic $D\to Pe^+\nu_e$ decays is written as
\begin{equation}\label{eq:ff_BK}
    f_+(q^2) = \frac{f_+(0)}{(1-q^2/m_{D^{*+}_{(s)}}^2)(1-\alpha q^2/m_{D^{*+}_{(s)}}^2)},
\end{equation}
where $m_{D^{*+}_{(s)}}$ is the mass of the $D^{*+}_{(s)}$ meson,
and $\alpha$ is a free parameter to be fitted.

The \textit{ISGW2 model}~\cite{ISGW2} assumes
\begin{equation}\label{eq:ff_ISGW2}
    f_+(q^2) = f_+(q^2_{\rm max}) \left( 1+\frac{r^2}{12}(q^2_{\rm max} - q^2) \right)^{-2},
\end{equation}
where $q^2_{\rm max}$ is the kinematical limit of $q^2$,
and $r$ is the conventional radius of the meson.

The most general parameterization of the form factor is the \textit{series expansion}~\cite{ff_zexpansion}, which is based on analyticity and unitarity.
In this parameterization, the variable $q^2$ is mapped to a new variable $z$ through
\begin{equation}
   z(q^2,t_0) = \frac{\sqrt{t_+-q^2}-\sqrt{t_+-t_0}}{\sqrt{t_+-q^2}+\sqrt{t_+-t_0}},
\end{equation}
with $t_{\pm}=(m_{D^+}\pm m_P)^2$ and $t_0 = t_+(1-\sqrt{1-t_-/t_+})$.
The form factor is then expressed in terms of the new variable $z$ as
\begin{equation}\label{eq:ff_series}
   f_+(q^2) = \frac{1}{P(q^2)\phi(q^2,t_0)} \sum_{k=0}^{\infty} a_k(t_0)[z(q^2,t_0)]^k,
\end{equation}
where $a_k(t_0)$ are real coefficients.
The function $P(q^2)$ is $P(q^2) = z(t,m^2_{D^*_s})$ for $D\to K$ and $P(q^2)=1$ for $D\to \pi$.
The standard choice of $\phi(q^2,t_0)$ is
\begin{eqnarray}
  \phi(q^2,t_0) &=& \left( \frac{\pi m^2_c}{3} \right)^{1/2} \left( \frac{z(q^2,0)}{-q^2} \right)^{5/2} \left( \frac{z(q^2,t_0)}{t_0-q^2} \right)^{-1/2}
  \nonumber
  \\
  &\times& \left( \frac{z(q^2,t_-)}{t_--q^2} \right)^{-3/4} \frac{(t_+-q^2)}{(t_+-t_0)^{1/4}},
\end{eqnarray}
where $m_c$ is the mass of the charm quark.

In practical use, one usually makes a truncation of the above series.
After optimizing the form factor parameters, we obtain
\begin{equation}\label{eq:ff_3series}
    f_+(q^2) =  \frac{f_+(0)P(0)\phi(0,t_0) (1+\sum_{k=1}^{k_{\rm max}}r_k [z(q^2,t_0)]^k)}{P(q^2)\phi(q^2,t_0) (1+\sum_{k=1}^{k_{\rm max}}r_k [z(0,t_0)]^k)},
\end{equation}
where $r_k\equiv a_k(t_0)/a_0(t_0)$.
In this analysis we fit the measured decay rates to the two- or three-parameter series expansion,
\emph{i.e.}, we take $k_{\rm max}=1$ or $2$.
In fact, the $z$ expansion with only a linear term is sufficient to describe the data.
Therefore we take the two-parameter series expansion as the
nominal parameterization to determine $f_+^{K(\pi)}(0)$ and $|V_{cs(d)}|$.

\subsection{Fitting partial decay rates to extract form factors}

In order to determine the form factor parameters,
we fit the theoretical parameterizations to the measured partial decay rates.
Taking into account the correlations of the measured partial decay rates among $q^2$ bins,
the $\chi^2$ to be minimized in the fit is defined as
\begin{linenomath*}
\begin{equation}
    \chi^2 = \sum_{ij}
    (\Delta\Gamma_i-\Delta\Gamma_i^{\rm th})\mathcal C^{-1}_{ij}
    (\Delta\Gamma_j-\Delta\Gamma_j^{\rm th}),
\end{equation}
\end{linenomath*}
where $\Delta\Gamma_i$ is the measured partial decay rate in the $i$th $q^2$ bin,
$\mathcal C_{ij}^{-1}$ is the inverse matrix of the covariance matrix
$\mathcal C_{ij}$.
In the $i$th $q^2$ bin, the theoretical expectation of the partial decay rate is
obtained by integrating Eq.~(\ref{eq:dG_dq2}),
\begin{linenomath*}
\begin{equation}
  \Delta\Gamma_i^{\rm th}
  = \int_{q^{2}_{{\rm min},i}}^{q^{2}_{{\rm max},i}}
    X\frac{G^2_F}{24\pi^3}|V_{cs(d)}|^2p^3|f_+(q^2)|^2 dq^2,
\end{equation}
\end{linenomath*}
where $q^2_{{\rm min},i}$ and $q^2_{{\rm max},i}$ are the lower
and upper boundaries of that $q^2$ bin, respectively.

In the fits, all parameters of the form factor parameterizations are left free.
The central values of the form factor parameters are taken from the results obtained by fitting the data
with the combined statistical and systematic covariance matrix together.
The quadratic difference between the uncertainties of the fit parameters
obtained from the fits with the combined covariance matrix
and
the uncertainties of the fit parameters
obtained from the fits with the statistical covariance matrix only
is taken as the systematic error of the measured form factor parameter.
The results of these fits are summarized
in Table~\ref{tab:FF},
where the first errors are statistical and the second systematic.
 \begin{table*}
  \centering
   \caption{Summary of results of form factor fits for $D^+\to \bar K^0e^+\nu_e$ and $D^+\to \pi^0e^+\nu_e$, where the first errors are statistical and the second systematic.}
   \label{tab:FF}
   \begin{ruledtabular}
   \begin{tabular}{lccc}
    \multicolumn{4}{c}{\it Single pole model}  \ST \\
    Decay mode & $f_{+}(0)|V_{cq}|$ & $m_{\rm pole}$ (GeV$/c^2$) \\
    $D^+\to \bar K^0 e^+\nu_e$ & $0.7094\pm0.0035\pm 0.0111$ & $1.935\pm0.017\pm 0.006$ \\
    $D^+\to \pi^0 e^+\nu_e$    & $0.1429\pm0.0020\pm 0.0009$ & $1.898\pm0.020\pm 0.003$ \\
    \hline
    \multicolumn{4}{c}{\it Modified pole model} \ST \\
    Decay mode & $f_{+}(0)|V_{cq}|$ & $\alpha$ \\
    $D^+\to \bar K^0 e^+\nu_e$ & $0.7052\pm0.0038\pm 0.0112$ & $0.294\pm0.031\pm 0.010$ \\
    $D^+\to \pi^0 e^+\nu_e$    & $0.1400\pm0.0024\pm 0.0010$ & $  0.285\pm0.057\pm 0.010$ \\
    \hline
    \multicolumn{4}{c}{\it ISGW2 model} \ST \\
    Decay mode & $f_{+}(0)|V_{cq}|$ & $r $ (GeV$^{-1}c^2$) \\
    $D^+\to \bar K^0 e^+\nu_e$ & $0.7039\pm0.0037\pm 0.0111$ & $  1.587\pm0.023\pm 0.007$ \\
    $D^+\to \pi^0 e^+\nu_e$    & $0.1381\pm0.0023\pm 0.0007$ & $  2.078\pm0.067\pm 0.011$ \\
    \hline
    \multicolumn{4}{c}{\it Two-parameter series expansion} \ST \\
    Decay mode & $f_{+}(0)|V_{cq}|$ & $r_1$ \\
    $D^+\to \bar K^0 e^+\nu_e$ & $0.7053\pm0.0040\pm 0.0112$ & $-2.18\pm0.14\pm 0.05$ \\
    $D^+\to \pi^0 e^+\nu_e$    & $0.1400\pm0.0026\pm 0.0007$ & $-2.01\pm0.13\pm 0.02$ \\
    \hline
    \multicolumn{4}{c}{\it Three-parameter series expansion} \ST \\
    Decay mode & $f_{+}(0)|V_{cq}|$ & $r_1$ & $r_2$\\
    $D^+\to \bar K^0 e^+\nu_e$ & $0.6983\pm0.0056\pm 0.0112$ & $ -1.76\pm0.25\pm 0.06$ & $-13.4\pm6.3\pm 1.4$ \\
    $D^+\to \pi^0 e^+\nu_e$    & $0.1413\pm0.0035\pm 0.0012$ & $ -2.23\pm0.42\pm 0.06$ & $  1.4\pm2.5\pm 0.4$ \\
   \end{tabular}
   \end{ruledtabular}
 \end{table*}

\begin{figure*}
\centerline{
\includegraphics[width=0.42\textwidth]{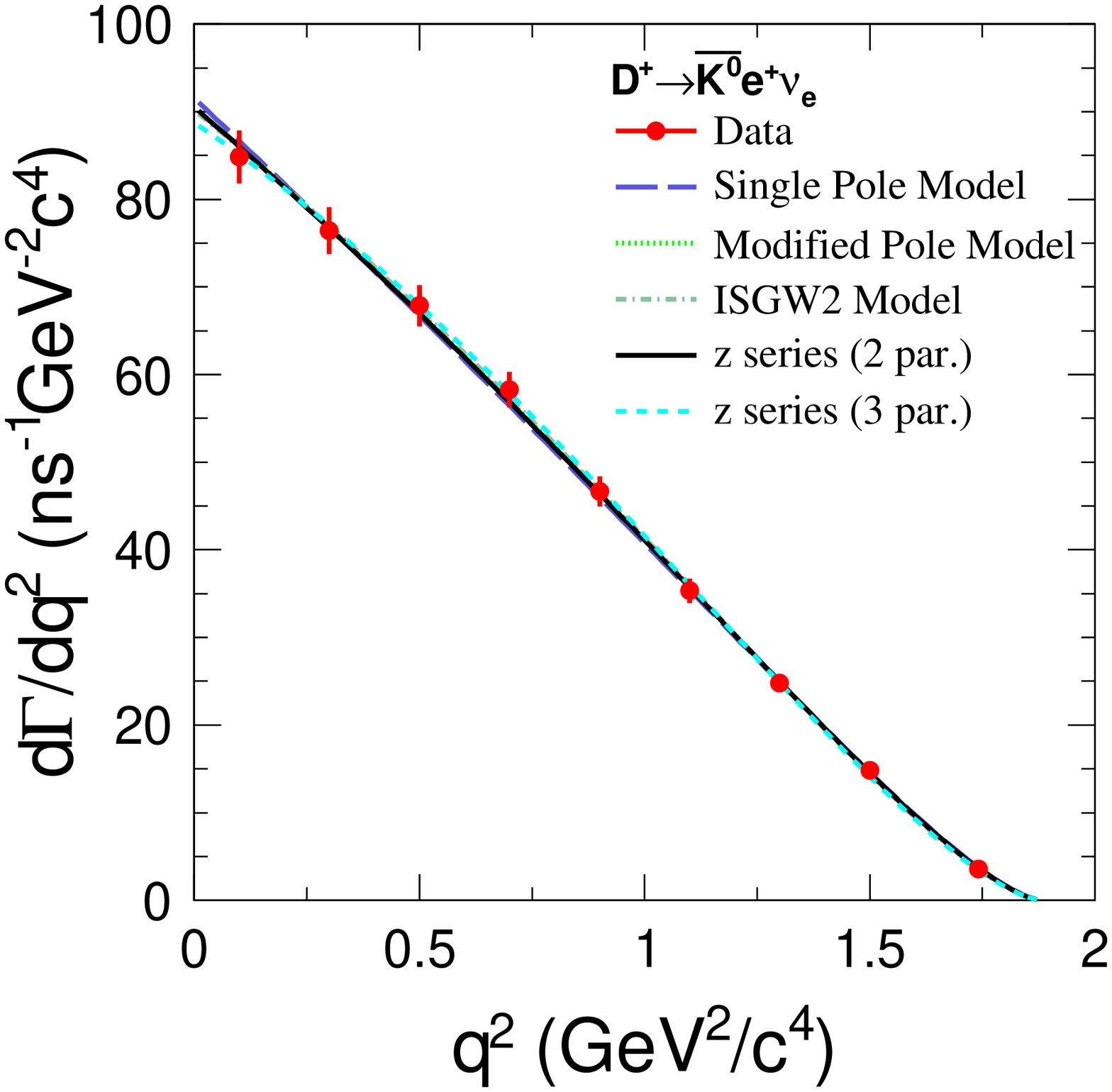}
\includegraphics[width=0.42\textwidth]{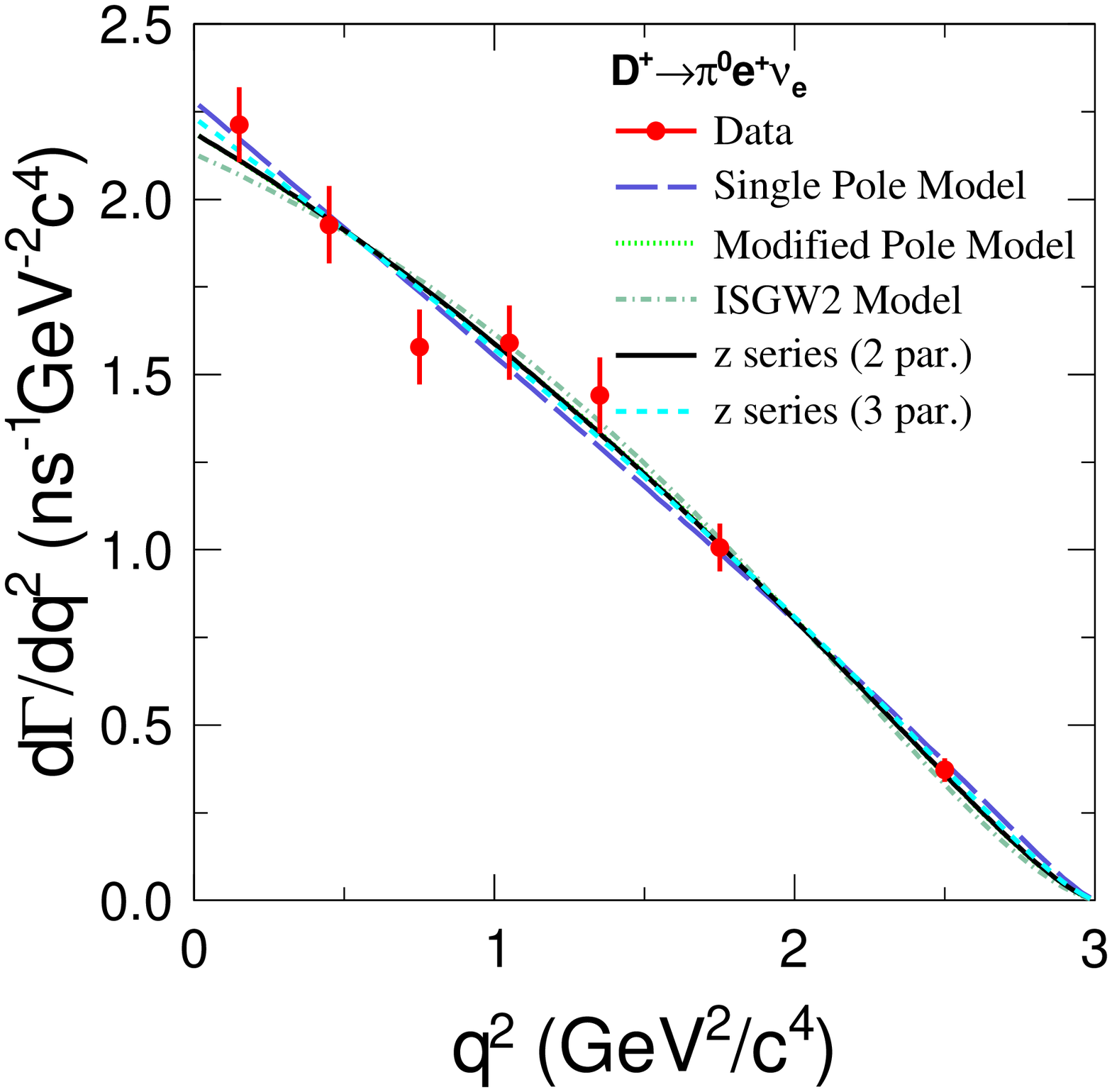}
}
\caption{
Differential decay rates for $D^+\to \bar K^0 e^+\nu_e$ (left)
and $D^+\to \pi^0 e^+\nu_e$ (right) as a function of $q^2$.
The dots with error bars show the data
and the lines give the best fits to the data
with different form factor parameterizations.
}
\label{fig:DR}
\end{figure*}

\begin{figure*}
\centerline{
\includegraphics[width=0.42\textwidth]{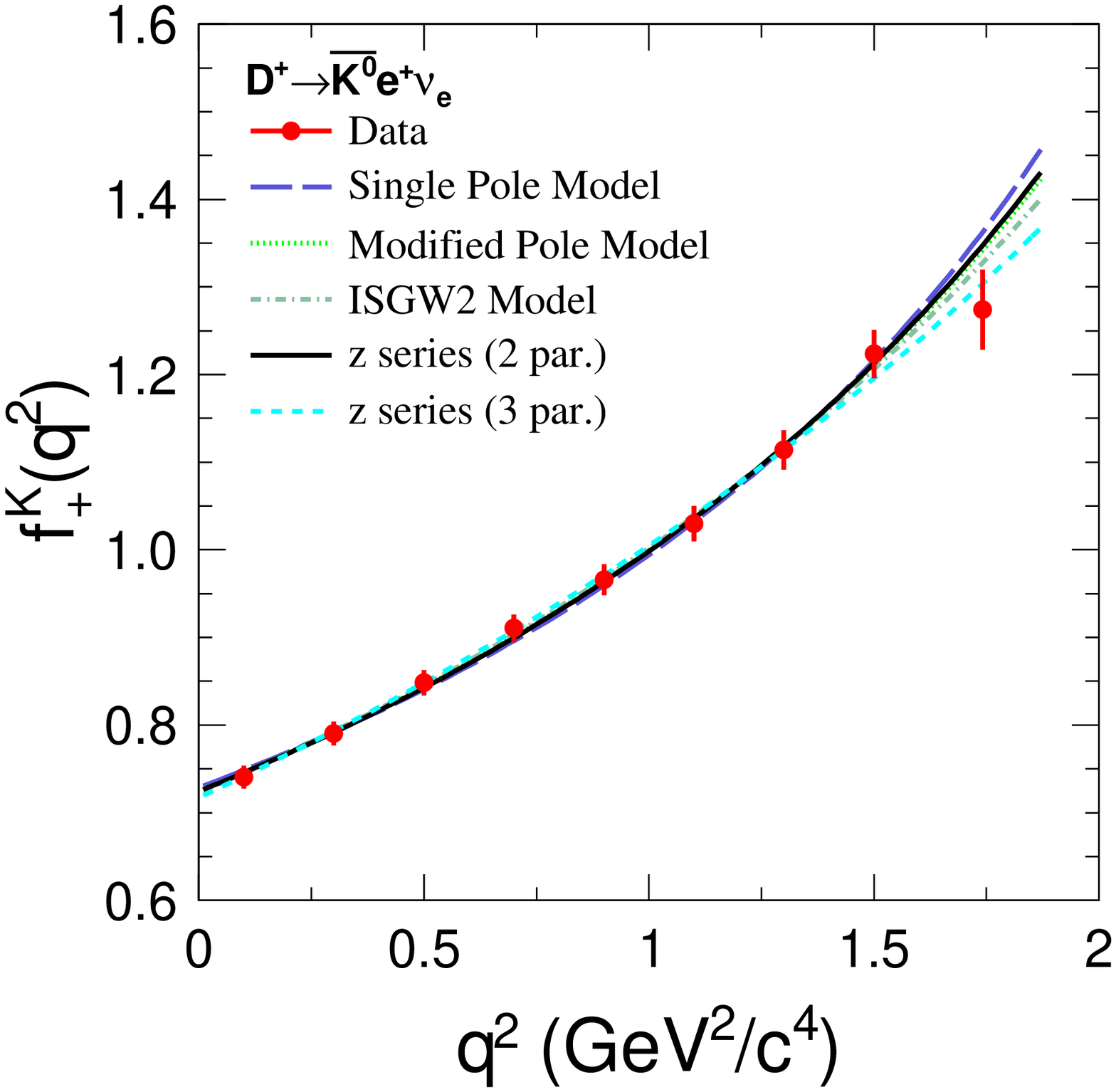}
\includegraphics[width=0.42\textwidth]{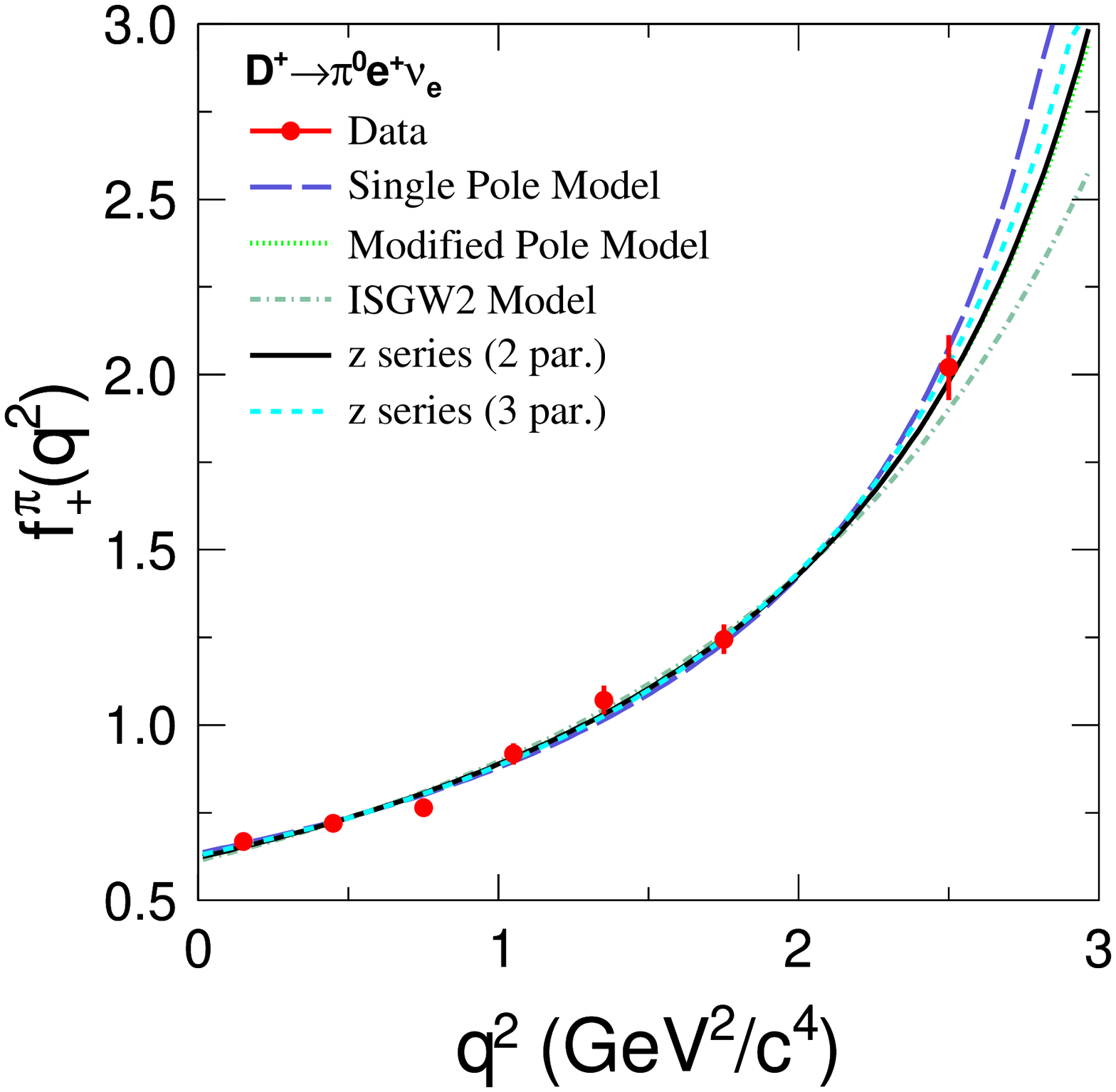}
}
\caption{
Projections on $f_+(q^2)$  for $D^+\to \bar K^0 e^+\nu_e$ (left)
and $D^+\to \pi^0 e^+\nu_e$ (right) as function of $q^2$,
where the dots with error bars show the data
and the lines give the best fits to the data
with different form factor parameterizations.
}
\label{fig:FF}
\end{figure*}

Figure~\ref{fig:DR} shows the fits to the measured differential decay rates
for $D^+\to \bar K^0 e^+\nu_e$ and $D^+\to \pi^0 e^+\nu_e$.
Figure~\ref{fig:FF} shows the projection of
the fits onto $f_+(q^2)$ for the $D^+\to\bar K^0e^+\nu_e$ and $D^+\to \pi^0 e^+\nu_e$
decays, respectively.
In these two figures, the dots with error bars show the measured values of
the form factors, $f_+(q^2)$, in the center of each $q^2$ bin, which are obtained with
\begin{linenomath*}
\begin{equation}
f_+(q^2_i)=\sqrt{ \frac{\Delta\Gamma_i}{\Delta q^2_i}
\frac{24\pi^3} {XG_F^2 {p^\prime_i}^3 |V_{cq}|^2} }
\end{equation}
\end{linenomath*}
in which
\begin{linenomath*}
\begin{equation}
{p^\prime_i}^3 = \frac{ \int_{q^2_{{\rm min},i}}^{q^2_{{\rm max},i}} p^3|f_+(q^2)|^2 dq^2}
{ |f_+(q^2_i) |^2 (q^2_{{\rm max},i}-q^2_{{\rm min},i}) },
\end{equation}
\end{linenomath*}
where
$|V_{cs}|=0.97351\pm 0.00013$ and $|V_{cd}|=0.22492\pm 0.00050$
are taken from the SM constraint fit~\cite{pdg2016}.
In the calculation of ${p^\prime_i}^3$, $f_+(q^2)$ is computed
using the two parameter series parameterization with the measured parameters.

\subsection{Determinations of $f_+^K(0)$ and $f_+^\pi(0)$}
Using the $f_+^{K(\pi)}(0)|V_{cs(d)}|$ values from the two-parameter
series expansion fits and
taking the values of $|V_{cs(d)}|$ from the SM constraint fit~\cite{pdg2016} as inputs,
we obtain the form factors
\begin{linenomath*}
\begin{equation}
f_+^{K}(0)=0.725\pm0.004\pm 0.012
\end{equation}
\end{linenomath*}
and
\begin{linenomath*}
\begin{equation}
f_+^{\pi}(0)=0.622\pm0.012\pm 0.003,
\end{equation}
\end{linenomath*}
where the first errors are statistical and the second systematic.

\section{Determinations of $|V_{cs}|$ and $|V_{cd}|$}
\label{sec:CKM}

Using the values of $f_+^{K(\pi)}(0)|V_{cs(d)}|$ from the two-parameter
$z$-series expansion fits
and in conjunction with the form factor values
$f_+^{K}(0)=0.747 \pm 0.011 \pm 0.015$~\cite{LQCD_fK}
and $f_+^{\pi}(0)=0.666 \pm 0.020 \pm 0.021$~\cite{LQCD_fpi} calculated from LQCD,
we obtain
\begin{linenomath*}
\begin{equation}
    |V_{cs}|=0.944 \pm 0.005 \pm 0.015 \pm 0.024
\end{equation}
\end{linenomath*}
and
\begin{linenomath*}
\begin{equation}
    |V_{cd}|=0.210 \pm 0.004 \pm 0.001 \pm 0.009,
\end{equation}
\end{linenomath*}
where the first uncertainties are statistical,
the second systematic,
and the third are due to the theoretical uncertainties in
the LQCD calculations of the form factors.

\section{Summary}
\label{sec:sum}

In summary, by analyzing 2.93~fb$^{-1}$ of data collected at 3.773~GeV with the BESIII detector at the BEPCII,
the semileptonic decays for $D^+\to\bar K^0e^+\nu_e$ and $D^+\to\pi^0e^+\nu_e$ have been studied.
From a total of $1703054 \pm 3405$ $D^-$ tags, $26008\pm 168$ $D^+ \to \bar K^0e^+\nu _e$ and
$3402\pm 70$ $D^+ \to \pi^0e^+\nu_e$ signal events are observed in the system recoiling against
the $D^-$ tags. These yield the absolute decay branching fractions to be
$\mathcal B(D^+ \to \bar K^0e^+\nu_e)=(8.60\pm 0.06 \pm 0.15)\times10^{-2}$ and
$\mathcal B(D^+ \to \pi^0e^+\nu_e)=(3.63\pm 0.08\pm 0.05)\times10^{-3}$.

We also study the relations between the partial decay rates and squared 4-momentum transfer $q^2$ for these two decays
and obtain the parameters of different
form factor parameterizations.
The products of the form factors and the related CKM matrix elements
extracted from the two-parameter series expansion parameterization
are selected as our primary results.  We obtain
$f_{+}(0)|V_{cs}| = 0.7053\pm0.0040\pm 0.0112$ and
$f_{+}(0)|V_{cd}| = 0.1400\pm0.0026\pm 0.0007$.
Using the global SM fit values for $|V_{cs}|$ and $|V_{cd}|$, we obtain the form factors
$f^K_+(0) = 0.725\pm0.004\pm 0.012$ and
$f^{\pi}_+(0) = 0.622\pm0.012\pm 0.003$.
Furthermore, using the form factors predicted by the LQCD
calculations, we obtain the CKM matrix elements
$|V_{cs}|=0.944 \pm 0.005 \pm 0.015 \pm 0.024$ and
$|V_{cd}|=0.210 \pm 0.004 \pm 0.001 \pm 0.009$,
where the third errors are dominated by the theoretical
uncertainties in the LQCD calculations of the form factors.

\begin{acknowledgments}
The BESIII collaboration thanks the staff of BEPCII and the IHEP computing center for their strong support. This work is supported in part by National Key Basic Research Program of China under Contract Nos. 2009CB825204, 2015CB856700; National Natural Science Foundation of China (NSFC) under Contracts Nos. 10935007, 11235011, 11305180, 11322544, 11335008, 11425524, 11635010; the Chinese Academy of Sciences (CAS) Large-Scale Scientific Facility Program; the CAS Center for Excellence in Particle Physics (CCEPP); the Collaborative Innovation Center for Particles and Interactions (CICPI); Joint Large-Scale Scientific Facility Funds of the NSFC and CAS under Contracts Nos. U1232201, U1332201, U1532257, U1532258; CAS under Contracts Nos. KJCX2-YW-N29, KJCX2-YW-N45; 100 Talents Program of CAS; National 1000 Talents Program of China; INPAC and Shanghai Key Laboratory for Particle Physics and Cosmology; German Research Foundation DFG under Contracts Nos. Collaborative Research Center CRC 1044, FOR 2359; Istituto Nazionale di Fisica Nucleare, Italy; Koninklijke Nederlandse Akademie van Wetenschappen (KNAW) under Contract No. 530-4CDP03; Ministry of Development of Turkey under Contract No. DPT2006K-120470; The Swedish Resarch Council; U. S. Department of Energy under Contracts Nos. DE-FG02-05ER41374, DE-SC-0010118, DE-SC-0010504, DE-SC-0012069; U.S. National Science Foundation; University of Groningen (RuG) and the Helmholtzzentrum fuer Schwerionenforschung GmbH (GSI), Darmstadt; WCU Program of National Research Foundation of Korea under Contract No. R32-2008-000-10155-0.
\end{acknowledgments}

\end{document}